# *In vitro* to *in vivo* acetaminophen hepatotoxicity extrapolation using classical schemes, pharmaco-dynamic models and a multiscale spatial-temporal liver twin

**Jules Dichamp[1,&], Geraldine Cellière[1,&], Ahmed Ghallab[2,3,&], Reham Hassan[2,3], Noemie Boissier[1], Ute Hofmann[4], Joerg Reinders[2], Selahaddin Sezgin[5], Sebastian Zühlke[6], Hengstler[2], Drasdo[1,2]**

[1]Group SIMBIOTX, INRIA Saclay-Île-de-France, 1 Rue Honoré d'Estienne d'Orves, 91120 Palaiseau, France

[2]Leibniz Research Centre for Working Environment and Human Factors, Technical University Dortmund, Ardeystr. 67, 44139, Dortmund, Germany.

[3]Department of Forensic Medicine and Toxicology, Faculty of Veterinary Medicine, South Valley University, 83523, Qena, Egypt.

[4]Dr. Margarete Fischer-Bosch Institute of Clinical Pharmacology and University of Tübingen, Auerbachstr. 112, 70376 Stuttgart, Germany.

[5]Faculty of Chemistry and Chemical Biology, TU Dortmund, Dortmund, Germany.

[6]Center for Mass Spectrometry (CMS), Faculty of Chemistry and Chemical Biology, TU Dortmund University, Dortmund, Germany.

[&]Shared first authors

**\* Correspondence:**
Corresponding Author
dirk.drasdo@inria.fr

**Abstract**

*In vitro* to *in vivo* extrapolation represents a critical challenge in toxicology. In this paper we explore extrapolation strategies for acetaminophen (APAP) based on mechanistic models, comparing classical (CL) homogeneous compartment pharmaco-dynamic (PD) models and a spatial-temporal (ST), multiscale digital twin model resolving liver microarchitecture at cellular resolution. The models integrate consensus detoxification reactions in each individual hepatocyte. We study the consequences of the two model types on the extrapolation and show in which cases these models perform better than the classical extrapolation strategy that is based either on the maximal drug concentration (Cmax) or the area under the pharmaco-kinetic curve (AUC) of the drug blood concentration. We find that an CL-model based on a well-mixed blood compartment is sufficient to correctly predict the *in vivo* toxicity from *in vitro* data. However, the ST-model that integrates more experimental information requires a change of at least one parameter to obtain the same prediction, indicating that spatial compartmentalization may indeed be an important factor.

# 1 INTRODUCTION

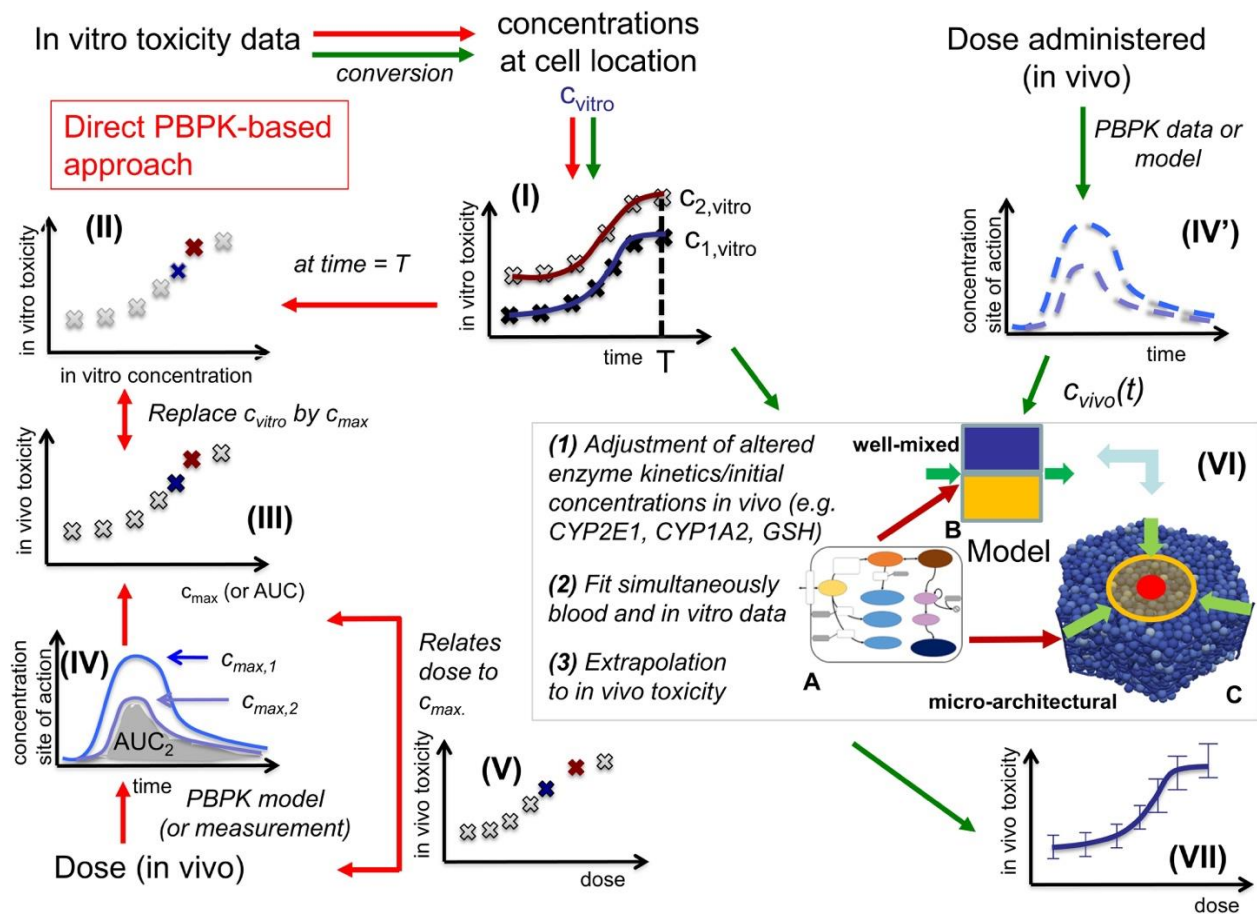

**Figure 1:** *In vitro - in vivo* **extrapolation approaches**. *In vitro*, toxicity is measured by exposing cell populations of interest to certain concentrations $C_{vitro}$ of drug (here: APAP) and measuring the fraction of cell death with time (I). The toxicity at a certain time T (often: 24h, or 3d) is considered as the time at which the toxicity is compared (II). A common consensus for extrapolation to *in vivo* is that the *in vitro* concentration value $C_{vitro}$ is identified by the $C_{max}$-value (III) of the corresponding pharmaco-kinetic (PK) curve for the drug in the blood (IV). The PK curve depends on the administered drug dose, so that by identifying $C_{max}$ and $C_{vitro}$, the dose value associated with that $C_{max}$-value can be associated with the toxicity value associated with the corresponding $C_{vitro}$ value (V). (Alternatively, to the $C_{vitro}$ / $C_{max}$ values, the area under the curve (AUC) is used). The model-based strategy mimics drug toxicity in a model by simulating the process of drug detoxification *in vitro* by the cells, whereby the toxicity pathway is explicitly represented in each cell and eventually integrated with a PK model (IV) and a compartment model of the organs of interest (VI). The latter may represent organ microarchitecture, here the liver lobule (VI-C) as repetitive minimal tissue unit or consider a well-mixed compartment (VI-B), both integrating an intracellular PD model (VI-A). Simulations with (VI-B, C) directly yield the *in vivo* toxicity (VII).

As direct risk-assessment of chemicals in human is, of course, prohibited, risk-evaluation usually is based on animal experiments. For cosmetics, animal experiments are forbidden in EU since 03/2013[1]. Large ongoing research programs focus on the ultimate goal to deliver testing strategies that enable animal-free risk assessment of chemicals (Godoy et al., 2013; Albrecht et al., 2019; Cherianidou et al., 2022). However, despite of large efforts in quantitative risk evaluation by *in vitro* and/or *in silico* methods, identification of no-observed-adverse-effect levels (NOAEL) or lowest-observed-adverse-effect levels (LOAEL) by alternative methods is presently not possible with satisfactory accuracy (Leist et al., 2017). Therefore, replacement of animal experiments has so far not been possible in most fields of regulatory toxicology. However, long-term, extrapolation of adverse drug effects to humans from *in vitro* experiments constitutes a key objective (Brecklinghaus et al., 2022).

*In vitro* testing usually begins with the establishment of a concentration-response relationship in cell cultures or in more complex 3D culture systems (Fig. 1(I)). Typical responses studied in relation to concentration of a test compound are cytotoxicity or biomarkers linked to adverse effects. A next challenge is *in vitro* to in *vivo* extrapolation (Sachinidis et al., 2019). Here, an *in vitro* concentration-response curve is usually translated to a dose-response curve, e.g. of organ toxicity, *in vivo* (Fig 1(V)). The easiest way of *in vitro - in vivo* extrapolation is to compare toxicity at *in vivo* relevant concentration ranges. An accepted procedure is the quantitative comparison based on the plasma peak concentration (Cmax ; $C_{max}$) or the area under the curve (AUC) of a test compound (Yu and Li, 2014) (Fig. 1(IV), Fig. 1(III)); named "direct Physiologically-Based-Pharmacokinetic (PBPK) based extrapolation'' from now on. This procedure often fails because: (1) *In vitro*, concentrations of test compounds in the culture medium are either relatively constant during the exposure period or decrease only slowly, due to the much larger volume of culture medium compared to the cell fraction. In contrast, faster concentrations changes occur *in vivo*, where compounds are eliminated by zero or first order kinetics unless the compound is infused or delivered over a long period of time, which is impractical for many drugs. (2) *In vitro / in vivo*-differences in the enzyme activities; and (3) the spatial organization of cells impacting the spatial-temporal transport and detoxification profiles.

In this work, we explore *in vitro* to *in vivo* extrapolation based on mechanistic computational models (CM) for the hepatotoxicity of APAP, taking the above three factors (1)-(3) into account. We directly compare to experimental *in vitro* and *in vivo* data. We start with a simple model and a simple parameter identification strategy and increase subsequently the complexity of both until the final model is able to explain the experimental data and to predict *in vivo* from *in vitro* hepatotoxicity.

APAP is a well-suited, clinically relevant toxic model substance to study *in vitro – in vivo* extrapolation strategies. Indeed, APAP overdose is the main reason for acute liver failure in several countries (Olson et al., 2017). An APAP overdose is cytotoxic for hepatocytes expressing cytochrome P450 (CYP) enzymes; mainly Cyp2e1, Cyp1a2 and Cyp3a4 (McGill and Jaeschke, 2013). These are localized close to the central vein of the liver lobules (Ghallab et al., 2019b), which constitute the repetitive anatomical and functional units of the liver. Liver lobules have a complex architecture facilitating exchange of metabolites between blood and hepatocytes, the parenchyma of the liver. Some chemical reactions are only executed in specific spatial regions of a lobule and the reaction rates vary according to a gradient in the periportal-pericentral axis, a phenomenon called zonation (Gebhardt, 1992; Bartl et al., 2015). In animal models, here mice, it is possible to dose-dependently determine the fraction of dead hepatocytes induced by a test compound *in vivo*; this can be done under conditions where blood is sampled time-dependently from relevant sites, e.g. the portal vein as liver inflow, the liver vein as liver outflow or mixed venous blood, as internal exposure monitoring (Sezgin et al., 2018). Moreover, hepatocytes can be isolated from the same mouse strain and incubated concentration-dependently *in vitro* (Ghallab et al., 2022).

---

[1] https://single-market-economy.ec.europa.eu/sectors/cosmetics/ban-animal-testing_en

Because of its relevance, APAP hepatotoxicity has been modelled by numerous mathematical approaches (Ben-Shachar et al., 2012; Diaz Ochoa et al., 2012; Leclerc et al., 2015; Reddyhoff et al., 2015; Smith et al., 2016; Fu et al., 2018; Franiatte et al., 2019; Sridharan et al., 2021; Heldring et al., 2022). A common computational model approach for drug effects is Physiologically-based (PB) / pharmacokinetic (PK) / pharmacodynamic (PD) models, that mimic either together or separately the concentration changes of the administered drug with time (PK) and its impact on components of interest (cells, tissue, body) (PD), taking often the physiology into account (PB) (Meyer et al., 2012; Upton and Mould, 2014; Kuepfer et al., 2016).

Here we directly confront two types of CMs, firstly a classical PB/PK/PD model (CL-model) approach, were we assumed for the liver a well-mixed compartment composed of CYP-positive-and CYP-negative hepatocytes (indicated by the yellow/blue box in Fig. 1(VI-B)), and secondly a spatial-temporal PB/PK/PD-model (ST-model) resolving liver microarchitecture (indicated by the lobule in Fig. 1(VI-C)), with experimental *in vitro* and *in vivo* mouse data on APAP-based hepatotoxicity. The "well-mixed" model approximates the population of hepatocytes as a stirred container filled with hepatocytes. I.e., the CL-model for the *in vivo* situation does not distinguish spatially separated compartments to reflect the spatial zonation of enzymes inside a liver lobule. The ST-model represents each individual hepatocyte as basic model unit inside a realistic sinusoidal network (Hoehme et al., 2010; Drasdo et al., 2014) representing a digital twin model of APAP toxicity in liver microarchitecture, hence naturally captures the effect of liver zonation (which is not the case for the well-mixed model). The APAP detoxification model is executed inside each individual hepatocyte in both, the well-mixed and the ST-model (Fig. 1(VI-A)).

We tested several model variants for the CL-model and different model-based extrapolation strategies (summarized in SFig. 1). Many a-priori plausible strategies did not function. The final strategy provided us with a parameter set that gives a valid prediction. However, the parameter identification procedure in that strategy (explained below) provides several – different - parameter sets leading to almost the same agreement to the experimental data sets that were used for the parameter identification procedure, but not all parameter sets lead to an *in vivo* toxicity prediction within the experimental uncertainty. The failures are informative and indicate that limitation of parameter ranges by proper experiments is indispensable for reproducibility of a model-based extrapolation. For successful parameter sets the extrapolation outperforms the classical extrapolation strategy based on Cmax and AUC, which we performed as reference (Fig. 1(IV)). The difference found between the CL-model and ST-model for our simulations indicate that a ST-model may be required for accurate simulations of APAP detoxification. This aside, an advantage of the ST-model is that it permits to directly include architectural distortions as for example fibrotic zones in liver fibrosis and cirrhosis, which is not possible in the CL-model.

The successful extrapolation simulation pursued in this work was composed of the following steps (Fig. 3 C.1 & C2, D.1 & D.2).

(1) As input data for the calibration of the *in vivo* model (CL-model or ST-model) hepatotoxicity data from the *in vitro* measurements (Fig. 1(I), Fig. 2D-F), and the APAP pharmaco-kinetics (Fig.1(IV'), Fig. 2I) were determined. More specifically, this data serves to calibrate the model parameters of a model that composed of an intracellular PD model coupled to a supra-cellular PK model composed of different body compartments (Fig. 3C.1).

(2) An intracellular PD model (Fig. 1(VI-A)) and an organ compartment (physiologically-based) PB/PK model were set up, whereby for the latter the liver was represented by two distinct approaches:

(3a) A well-mixed liver model whereby all cells were considered as independently fed by the APAP blood concentration (Fig. 1(VI-B); Fig. 3C.1). The model integrating the intracellular PD-model into that well-mixed liver model is a classical (CL-)model (denoted as CL-model 3 in Fig. 3).

(3b) A spatial-temporal (ST-) resolved liver micro-architectural multiscale model, where the intracellular PD model was executed in each individual cell of a virtual lobule (Fig. 1(VI-C)), referred to as ST-model (Fig. 3D.1).

(4) Within a single fit procedure (Fig. 3C.1), the parameters of the intracellular PD model and the parameters of the PB/PK model (together forming CL-model 3, cf. Fig. 3C.1) were fitted together subject to the following conditions: (i) the parameters of the intracellular PD models were fitted to the *in vitro* data (Fig. 1(I)). (ii) For the simultaneous fit of the (coupled) PD/PB/PK model (composed of intracellular PD-model and the well-mixed PB/PK model, cf. Fig. 3C.1) to the PK data (Fig. 1(IV')), the enzyme activities for CYP2E1 and CYP1A2 and the GSH levels obtained from the fit to the *in vitro* data (i) were corrected by the ratios found from an experimental determination of enzyme activities from *in vivo* hepatocytes. I.e., despite the fits to the *in vitro* toxicity data and to the *in vivo* PK-data were executed simultaneously, the parameters of the intracellular PD-model used in both sub-steps (*in vitro* / *in vivo*) were not the same but differed by the levels of the aforementioned three factors (CYP2E1, CYP1A2, GSH). However, the model structure of the intracellular PD-model as well as the other parameters were the same.

Once a suitable fit of the parameters of CL-model 3 was obtained, the so parameterized CL-model 3 composed of parameter sets {intracell-PD, PB/PK, cf. Fig. 3C.1} was executed to predict the *in vivo* toxicity (Fig. 3C.2), and the simulation results were compared to experimental data (Fig. 1(VII)).

(5) The same parameter set was then chosen to execute the spatial temporal (ST-)model (Fig. 3D.1), whereby the micro-architecturally resolved model required one additional parameter, the liver flow rate. The aim was to study, whether the ST-model would basically reproduce the results of the well-mixed model, or if its results would differ, indicating that APAP-gradients in the liver lobule may not be negligible.

We found that the CL-model and the ST-model performed similarly for the same intracellular parameters and the chosen lobule geometry and topology only, if the volume flow rate was in the lower range of values compatible with published references. However, a small volume flow rate required extracellular parameters to be different in the CL-model and the ST-model unless the blood flow speed was permitted to be markedly smaller than the velocity values found in published references. The extrapolated *in vivo* toxicity results partially performed very well. However, for the fit of the intracellular PD-model, different parameter combinations performed equally well with regard to the fitness criterion, but they did not with regard to the *in vivo* toxicity prediction. This indicates that the parameter landscape is rough so does not guarantee convergence to the same set of intracellular model parameters.

From the simulation results it would be expected that further experiments to narrow the parameter ranges of the intracellular model are likely to result reproducibly in valid *in vivo* extrapolations that outperform the classical extrapolation schemes based on Cmax and AUC, as this is the case for the specific parameter sets found below (Figs. 6, 7, 8). Such experiments may be complemented by determination of extracellular parameters such as the flux of APAP into the hepatocytes. However, such experiments combining intra-and extra-cellular levels are very laborious, and would thus call for a community effort. Such an effort may also imply the extension to human data.

Below, first an overview of the tested strategies and models is given, explaining which of the strategies failed, and in which way they failed, before finally explaining the model that succeeded. The success case confirms that a CL-model as well as a ST-model is a-priori capable to simultaneously explain *in vitro* and *in vivo* hepatotoxicity, so the structure is a priori suited. However, validation of the model would require to further constrain its parameters, which was out of the feasibility of this work. The ST-model is presented here as a proof of a concept of a spatially resolved multi-scale model of *in vivo* hepatotoxicity. Such models are expected to capture key components of hepatotoxicity (Holzhütter et al., 2012; Schwen et al., 2016; Ho and Zhang, 2020).

## 2    MATERIALS AND METHODS

### 2.1    Animal experiments

Eight-ten-week-old male C57Bl/6N mice, 20-25 grams body weight were used (Janvier labs, France). The mice were fed ad libitum with the Ssniff R/M-H, 10 mm standard diet (Ssniff, Soest, Germany) and housed at controlled ambient temperature of 25∘ C with 12 h day, 12 h night cycles. All experiments were approved by the local animal ethics committee.

### 2.2    Induction of liver injury by APAP.

In order to test the dose response of APAP in livers of mice, various doses (89 up to 500 mg/kg) were administered, where APAP was dissolved in warm PBS and injected intraperitoneally (i.p.) into overnight fasted mice, with an application volume of 20 ml/kg (Schneider et al., 2021b). On day one after APAP injection, liver tissue samples were collected and processed for histopathology and immunohistochemistry analyses as previously described (Campos et al., 2020; Holland et al., 2022) ). Three mice were used for each tested dose. All experiments with mice were approved by the local authorities.

### 2.3    Pharmacokinetic analysis of acetaminophen.

For pharmacokinetic analysis of APAP, a dose of 450 mg/kg APAP was administered i.p.. Blood samples were collected in a time-dependent manner after APAP injection (0, 5, 15, 30, 45, 60, 120, 240 and 480 minutes) from the portal vein 'representing 75% of the liver inflow', from the heart 'representing 25% of the liver inflow' and from the hepatic vein 'representing liver outflow' as described in Ghallab et al. (2016). After blood collection plasma was immediately separated by centrifugation at 13,000 rpm for 10 minutes and stored at -80℃ until analysis. APAP concentrations were determined by LC-MS/MS as previously described (Sezgin et al. 2018).

### 2.4    Hematoxylin and eosin staining

Hematoxylin and eosin (H&E) staining was performed in 5 µm-thick formalin-fixed paraffin-embedded liver tissue sections as described in Ghallab et al. (2021). Representative images were then acquired with a bright field microscope using Cell^F software (Olympus, Hamburg, Germany) and from these the dead cell areas determined.

## 2.5   CYP2E1 and CYP1A immunostaining

Immunostaining of CYP2E1 and CYP1A were performed in 5 µm-thick cryo-sections using antibodies against CYP2E1 (Cat. No. MFO-100, Stressgen, Victoria, BC, Canada) and CYP1A (a gift from Dr. R. Wolf, Biochemical Research Centre, University of Dundee, Dundee, UK). In order to detect antibody binding, the tissue sections were incubated with a horseradish peroxidase-conjugated secondary antibody (Cat. No. P0217, DakoCytomation Denmark A/S, Glostrop, Denmark). Following washing steps, antibody binding was visualized by covering the tissue sections with AEC+ high sensitivity substrate chromogen (Dako, USA) for 10 minutes. The AEC-stained sections were then preserved by mounting with an aqueous mounting media (Schenk et al., 2017). Quantification of the CYP positive area was done in whole slide scans as previously described (Ghallab et al., 2019b) (see also SI section 2.10).

## 2.6   Measurement of CYP450 enzyme activity in isolated liver microsomes

Liver microsomes were prepared as described before (Lang et al., 2001). Microsomal incubation mixtures contained 50 µg of microsomal protein in 0.1 M sodium phosphate buffer (pH 7.4), NADPH-generating system (5 mM $MgCl_2$, 4 mM glucose 6-phosphate, 0.5 mM NADP+, and 4.0 U/ml glucose 6-phosphate dehydrogenase), and 50 µM chlorzoxazone (CYP2E1 substrate) and phenacetin (CYP1A2 substrate) in a final volume of 100 µl. Samples were preincubated in a water bath for 5 min at 37 ºC and the reaction started by addition of the NADPH-generating system. Reactions were stopped after 15 min by adding 10 µl of 250 mM formic acid and 10 µl of internal standard solution containing a mixture of deuterium labelled analogues of the analytes and cooling on ice. The supernatant was analyzed by LC-MS/MS as described previously (Feidt et al., 2010).

## 2.7   Isolation and cultivation of primary mouse hepatocytes

Primary mouse hepatocytes were isolated from the livers of male C57Bl/6N mice, 8–10-week-old, (Janvier labs, France) as previously described (Godoy et al., 2016). Briefly, the liver was perfused through the vena cava for 15 minutes with EGTA buffer at 37°C. Subsequently, the liver was perfused with collagenase buffer for 5-7 minutes. The liver was then excised and dissociated in a suspension buffer. Following a filtration step, the cell suspension was centrifuged for 5 minutes at 50 g. The cells were re-suspended in 10 ml suspension buffer. The cell viability was checked by trypan blue exclusion. In order to prepare confluent hepatocyte culture, the cells were seeded in six-well plates at a density of 800,000 cells per well in Williams E medium supplemented with 10% fetal calf serum (FCS), 2mM L-glutamine, 100 units/ml penicillin, 0.1 mg/ml streptomycin, 10 µg/ml gentamycin (PAN Biotech, Aidenbach, Germany) and 100nM dexamethasone (Sigma-Aldrich, Munich, Germany). The cells were allowed to attach for 2h at 37°C, 5% CO2. Subsequently, the floating cells were washed out. In order to prepare sandwich cultures a second layer of collagen-1 was added and allowed to polymerize for 30 minutes. For all subsequent cultivations normal media was used without FCS. The constituents of the normal media are Williams E medium supplemented with 2mM L-glutamine, 100 units/ml penicillin, 0.1 mg/ml streptomycin and 10 µg/ml gentamycin. Hepatocyte viability was evaluated by trypan blue exclusion assay. Only when viability above 90% the cells were included in the experiments.

## 2.8   Measurement of CYP450 enzymes activity in cultivated hepatocytes

In order to measure the activities of CYP2E1 and CYP1A2 in isolated cells, primary mouse hepatocytes were incubated with 50 µM chlorzoxazone and phenacetin directly after isolation (fresh hepatocytes) or at different time intervals (0h, 2h, 4h, 1 day, 3 days and 7 days) after attachment of the cells. After 60 and 120 minutes of incubation, 50 µl of the cell culture supernatant was taken and the reaction was stopped by adding 5 µl of formic acid (250 mM). After mixing, the samples were stored at -20 °C until analysis. After thawing samples were spiked with internal standard mixture and analyzed by LC-MS/MS as described previously (Feidt et al., 2010).

## 2.9    Cytotoxicity of acetaminophen *in vitro*

To check APAP-induced cytotoxicity in cultivated primary mouse hepatocytes (sandwich cultures), the cells were incubated with various concentrations of APAP (0, 0.25, 0.5, 1, 2, 4, 8, 16 and 32 mM). Incubations periods started 2h after plating of hepatocytes. This early exposure was chosen because primary mouse hepatocytes are known to show a particularly fast decrease of expression of phase I and phase II metabolizing enzymes even when cultivated in organotypic 3D cultures (Godoy et al., 2016). 24 h later, the cells were incubated with propidium iodide (PI) 1.5 mM (Thermo Scientific, MA, USA) diluted 1:500 in cultivation media for 5 minutes at 37°C. Moreover, 3D cultivated mouse hepatocytes were incubated for 5 min or 3 h with APAP followed by washout and analysis of cell morphology and PI uptake at 24 h to understand the influence of the exposure period to APAP (SFig 1A, B). Already incubation for only 3h led to an increase in PI-positive hepatocytes compared to controls for all tested APAP concentrations (SFig 1C). PI positive nuclei were visualized using a combined fluorescence/phase contrast microscope (Nicon, Dusseldorf, Germany). Five representative images were quantified per well. The results are from three independent experiments (with different mice), and each experiment was performed in triplicate.

A second set of experiments was performed for the 4mM APAP-concentration at 5 different time points: 1, 4, 8, 18, 24 hours. This data set was rescaled to match the 24h data set (see SFig. 1). The scaling was necessary as the 24h-toxicity and the dead-cell kinetics were measured using hepatocytes from two different mice, each mouse having different levels of GSH prior to starvation that were not known.

## 2.10   Measurement of GSH concentration

The glutathione content in liver tissue homogenate was measured using LC-MS/MS as previously described (Sezgin et al., 2018).

## 2.11   Mathematical modeling

### 2.11.1   Pharmacokinetic (PK) model

The PK model considers two compartments: the peritoneal cavity into which APAP is injected, and the blood compartment, into which APAP diffuses from the peritoneal cavity. Clearance of APAP in the blood is modeled by a first-order decay term. Only a fraction of the dose injected intraperitoneally is assumed to reach the blood compartment according to a limited bio-availability. Parameters were fitted for the experimentally measured blood kinetics observed for the dose of 300 mg/kg of body weight and one data point measured at 24h each for 56 mg/kg, 167mg/kg, and 500mg/kg through minimization of the log-likelihood function using the CMA-ES algorithm (Hansen and Ostermeier, 1996; Hansen, 2006). In a next step, prediction of the concentration-time curve for the doses 89, 158, 281, 375 and 500 mg/kg was done in MATLAB using the estimated parameters. Equations are presented in the supplementary material. This model was used to perform the AUC and Cmax strategies, and for CL-model 1 (SFig. 1-A.2).

### 2.11.2  *In vitro*-PD compartment model of APAP

The *in vitro* metabolism of APAP was described using an ODE-based pharmacodynamic model. In brief, APAP diffuses passively through the cell membrane as this constitutes the main transport mechanism (Prescott, 1980). Once in the cell, it is metabolized by UGT and SULT to non-toxic metabolites and by CYP2E1 and CYP1A2, the two enzymes involved in APAP metabolism in mice (Snawder et al., 1994; Lee et al., 1996; Genter et al., 1998) to the toxic metabolite NAPQI. When the level of GSH, an intracellular antioxidant, is sufficiently high, NAPQI is detoxified by GST. If GSH levels get depleted through the detoxification reaction, NAPQI forms adducts in the mitochondrial proteins leading to the formation of ROS (reactive oxygen species), which get amplified through activation and translocation of JNK to the outer mitochondrial membrane. This also leads to further depletion of GSH. High levels of ROS trigger the MPT (mitochondrial permeability transition), leading to the impairment of ATP production. In our model, a cell is considered dead when its ATP level, initially around 5000 µM, falls below 100 µM. This value was chosen arbitrarily low relatively to the initial value (2% of the initial value) but selected simulations varying this parameter indicated that the parameter has only a negligible effect.

Given a cellular input exposure (concentration of APAP outside the cell over time) in the culture medium, the fraction of dead cells over time is obtained by simulating an ensemble of 3251 cells, each cell differing from the other one reflecting the cell-to-cell variability in the population (Furusawa et al., 2005; Sigal et al., 2006; Spencer et al., 2009). The population size of 3251 has been chosen to match it with the number of cells considered in the spatial temporal model of a virtual lobule (Hoehme et al., 2010) as detailed in section 2.11.4.

Cell-to-cell variability is modeled by varying all parameters that depend on enzyme levels (i.e. maximal velocities and production rates). The distribution of the varying parameters in the cell population is assumed to be log-normal around the mean parameter value, with a certain coefficient of variation CoVa (standard deviation over the mean). Except of the CYP2E1 and CYP1A2, which are zonated in the liver, the mean values and the CoVa are fit parameters that have been chosen to be the same for all values. The distribution of maximal velocities of CYP2E1 and CYP1A2 in the cell population has been chosen to reflect the experimentally measured spatial distribution *in vivo* (see SI section 1.10; for example, the fraction of cells with minimal CYP2E1-reaction velocity ($V_{max,CYP2E1}$, see SI) in the *in vitro* population corresponds to the fraction of cells with minimal CYP2E1-reaction velocity in the liver lobule).

Also, UGT, SULT and GST enzyme activities are zonated (Gebhardt, 1992) . The latter reference describes spatial gradients for these enzymes. It is possible to take these into account by modifying the associated reaction rates to reproduce their spatial distribution as has been done for example in (Means and Ho, 2019) in an extension of the metabolic model from (Reddyhoff et al., 2015). For our considered mouse model, there is no quantitative data available on those gradients, thus the gradients would become additional fit parameters. Given that the metabolic model is already complex, we avoided having such additional fit parameters. Thus, all mean values of reaction rates for sulfation, glucuronidation and glutathione-s-transferase are the same from one cell to another. It was verified that the results do not depend critically on the number of cells chosen. Using 160 cells instead of 3251 cells gave for the same seed an only very slightly different parameter distribution such that the fraction of death cells differs by only ~ ±5% from those where 3251 cells have been used.

The equations are presented in the supplementary material. They were implemented in MATLAB. Physiologically relevant parameter ranges were determined from literature for each parameter and parameters having an influence on the simulation were estimated using the *in vitro* fraction of dead cells after a 24h exposure to 0.25mM, 0.5mM, 1mM, 2mM, 4mM or 8mM APAP.

### 2.11.3 *In vivo*-PD compartment model of APAP

The *in vitro* PD model was modified in the following way to represent the *in vivo* settings:

i) The input APAP concentration outside the cells was replaced by the concentration of APAP in blood through two main strategies: (1) the pharmacokinetic model of APAP is plugged as an input concentration profile for 3215 cells representing a piece of a liver lobule (used in CL-model 1: Fig. 3(A.1, A2) and CL-model 2: Fig. 3(B1, B2) and (2) as obtained from a multi-compartment model (CL-model 3, Fig. 3(C.1)). The latter model considered a peritoneum compartment into which APAP was injected. From there, APAP is then released to a blood compartment from which the kidneys and liver eliminated it. The kidney effect was modeled by a first-order elimination term in the blood compartment. The liver uptake is modeled as the sum of the uptake by 3215 cells within a well-mixed liver compartment (Fig. 3(C1)), and later upscaled. The 3125 cells were chosen to match the cell population size of the spatial-temporal (ST-) model (Fig. 3(D.1) below. For upscaling, this sum is multiplied with the estimated number of equivalent liver lobule pieces, i.e. the ratio of the total number of cells in the liver and the number of cells in the spatial-temporal lobule (see supplementary material for more details, section).

ii) The CYP enzyme activities ($V_{max}$) were modified to capture the higher enzyme activity *in vivo* compared to the *in vitro* case (based on experimental data, see section 2.6 and 2.8 .

iii) The initial GSH concentration was modified to two times lower initial concentrations compared to the *in vitro* case (based on experimental data, see section 2.10 and SFig. 4).

Note that one could consider more refined compartment models representing certain zones by a well-mixed liver sub-compartments. For example, three well-mixed spatial compartments representing cell negative to both CYP2E1 and CYP1A2 enzymes, cells positive to CYP1A2 enzyme but negative to CYP2E1 enzyme and cells positive to both CYP2E1 and CYP1A2 enzymes. Connections between the compartments would then be considered by advective transport and thus the introduction of liver flow rate parameter which is present in the ST model (see SI section 1.6.4). We did not consider this approach here to reduce the complexity of the model but move directly from a single blood compartment that includes the hepatocytes to the ST-model.

### 2.11.4 *In vivo*-PD spatial-temporal model of APAP

The cells uptake in the space-free *in vivo*-PD model of APAP (sect. 2.10 above) was replaced by a spatial lobule of 3215 cells, representing each individual hepatocyte within liver lobule tissue microarchitecture. The detailed reconstruction of the liver lobule microarchitecture is described in (Hoehme et al., 2010), the experimental and image analysis protocols and procedures in (Hammad et. al., 2014), and the image processing software in (Friebel et. al. 2015). Briefly, confocal laser scans of up to 150μm of depth were used, stained for blood vessels, hepatic nuclei, and bile canaliculi permitting to process each of these components separately. First, the 3D sinusoidal network was reconstructed within an image processing pipeline of several filtering, segmentation and restoration steps (Friebel et. al. 2015). Then, the analysis of position and size of the hepatocytes was performed. From 26 liver lobule samples, a list of values for parameters characterizing the liver microarchitecture such as vessel diameter or density of the hepatocytes has been generated. Finally, a "statistical representative" liver lobule has been constructed by sampling from the parameter distributions.

The *in vivo*-PD model of APAP (from 2.10 above) was now solved in each hepatocyte. The spatial lobule represents a full lobule in (x, y)-direction with a height of 10 hepatocytes in z-direction. APAP enters the sinusoidal (liver capillary) network of the model liver lobule with the blood by the portal veins and hepatic artery, that are not distinguished in the model for the blood flow. The hepatocytes were assumed to take up APAP by the same mechanism as for the well-mixed liver model (sect. 1.11.3), whereby the blood concentration of APAP in the sinusoid at the position of each individual hepatocyte was used. The blood left the liver by the central vein hence representing the lobule outlet. The liver outlet is then connected back to the blood compartment. The entire liver compartment was modeled as parallel arrangement of identical liver lobules such that each sub-volume of blood entering the liver would pass only one liver lobule. Blood flow, transport in the blood vasculature and intra-cellular reactions were computed in only one lobule – that was considered as "representative" - whose contribution was multiplied by the total number of liver lobules. Because the liver is now considered as a spatial compartment, the coupling to the blood compartment involves the liver flow rate as an additional parameter. This parameter was calibrated manually around experimental data (see SI section 1.6.4).

For the detailed micro-architecture of the lobule in the computer simulation, statistically representative liver lobule architectures obtained from three-dimensional volume reconstructions of confocal laser scanning micrographs where chosen (Hoehme et al., 2010; Hammad et al., 2014). Simulations directly in the 3D volume reconstructions were not feasible as the 3D volume data sets did not represent the portal veins, while determination of the flow boundary conditions require an entire liver lobule in (x,y)-plane. The construction of representative liver lobules described in (Hoehme et al., 2010; Drasdo et al., 2014) circumvents this problem. It displays portal triads, the sinusoidal network, the hepatocytes, and the central vein.

Within the sinusoidal network, steady-state blood flow is assumed, and calculated in each sinusoid from the total entering flow via the portal vein and hepatic artery. Poiseuille-flow is considered, which relates the volume flow rate to the pressure difference along each sinusoid of the sinusoidal network. To account for the dependence of the effective blood viscosity with the sinusoid diameter (Fahraeus-Lindqvist effect) an empirical effective blood viscosity model was chosen (Boissier et al., 2021).

The transport of APAP with the blood flow, and its uptake by hepatocytes, is simulated by partial differential equations mimicking the advective transport of molecules within the blood, and the transport of APAP from the blood into and out of each hepatocyte. The *in vivo*-PD model is executed in each individual hepatocyte of the liver lobule with the hepatocytes' kinetic parameters varying due to cell-to-cell variability, now in space. (The equations and further details are provided in the supplementary material.)

### 2.11.5   Fitting procedure

A similar fitting procedure has been applied to the *in vitro* and *in vivo* data.

As the volume of the solution *in vitro* was large such the APAP concentration only changed negligibly over the measurement period of 24 hrs. Hence, simulation of the cytochrome P450-negative cells was unnecessary, and the fitting procedure could be focused on the CYP-positive cells only, which contained the NAPQI, SULT and the UGT conversion reactions.

To identify the values of the model parameters to the *in vitro* data, the following strategy was performed. First, the initial values of the 34 parameters and the ranges over which they can vary were chosen according to literature values. In the next step a simple parameter sensitivity on these 34 parameters was performed to identify those parameters, that influence the hepatotoxicity. Each parameter was separately varied by several orders of magnitude (divided and multiplied by 100) for one fixed concentration of APAP, while the other parameters were kept constant. From this analysis, 17 parameters were determined to only have a negligible impact on the hepatotoxicity, quantified by the simulated fraction of dead hepatocytes, and were thus fixed to their initial values. To calibrate the remaining 17 sensitive parameters, the log-likelihood function using all the available data of *in vitro* toxicity was maximized using the CMA-ES algorithm (a global convergence algorithm for non-linear functions). The finally computed dead cell fraction used the entire cell population size i.e., CYP-positive-and negative cells into account in the denominator. (More details and equations can be found in supplementary information.)

For the *in vivo* model fit, no sensitivity analysis to separate sensitive and non-sensitive parameters was performed as the number of parameters (5) was small, and the ranges were better known. However, as the APAP concentration varies markedly in the blood, now both, the cytochrome P450 positive and negative cells were taken into account, whereby the enzyme activities of the UGT and SULT-pathway were assumed to be the same in all hepatocytes.

## 3 RESULTS

### 3.1 Establishment of *in vitro* and *in vivo* data of APAP hepatotoxicity

In a first step we determined *in vitro* and *in vivo* APAP hepatotoxicity (corresponding to Fig. 1(I), (V)).

***In vitro:*** Primary hepatocytes were isolated from male C57BL6/N mice and cultivated in a 3D configuration between two soft gel collagen layers (SFig. 1). Six different concentrations ranging between 0.25 mM and 32 mM APAP were tested and cytotoxicity was analyzed by propidium iodide (PI) staining and cell morphology after a 24 h exposure period (Fig. 2A-C, SFig. 1C, D). For the 4mM dose, cytotoxicity was analyzed at 5 different time points (Fig. 2C). The cytotoxicity was measured in terms of the number of dead cells, that was manually quantified based on PI positive nuclei and cell morphology (SFig. 1B). The representative result of PI-stained hepatocytes demonstrates that the technique allows a clear differentiation between PI positive and negative hepatocytes (Fig 2B). The fraction of dead cells increases to almost its saturation value after 10 hours. Analysis of hepatocytes of three different mice resulted in a concentration-dependent increase of cytotoxicity, where approximately 40% of all hepatocytes lost their ability to exclude PI at concentrations of 4 mM and 8 mM APAP (Fig 2C). At even higher, concentrations (16 and 32 mM, not shown in Fig. 2C) a sharp increase of the concentration response curve was obtained, leading to cell death of all (i.e. not only CYP-positive) hepatocytes.

***In vivo:*** To determine the APAP-induced hepatotoxicity *in vivo* (Fig. 2D-F; SFig. 2), the same mouse strain (C57BL6/N, male) was studied that has been used to generate the *in vitro* data with cultivated hepatocytes. APAP was intraperitoneally administered at five doses ranging between 89 and 500 mg/kg body weight and livers were analyzed 24 h after administration (SFig 2A). The result demonstrates a dose-dependent increase of the pericentral dead cell area (SFig 2B). For doses up to 375 mg/kg the dead cell area corresponds approximately to the pericentral fraction of hepatocytes that express CYP2E1 (SFig 2C). Next, the pericentral dead cell area was quantified in relation to the total area of the liver tissue to calculate the fraction of dead cells *in vivo*. Mean values and standard deviations of three mice per dose showed a dose-dependent increase in the fraction of dead hepatocytes, with 89 mg APAP/kg b.w. representing the NOAEL and 158 mg APAP/kg b.w. the LOAEL (Fig 2F). These *in vivo*-hepatotoxicity data serve as reference for comparison to all *in vitro – in vivo* extrapolation strategies.

The here presented *in vivo* data of APAP are in agreement with previous reports on APAP-induced liver injury in mice (McGill et al., 2012; Ghallab et al., 2022). The *in vitro* experiments showed slightly higher APAP-induced cytotoxicity compared to previous reports (Jemnitz et al., 2008; Jaruchotikamol et al., 2009). A possible explanation of this discrepancy is that we started incubation with APAP already at two hours after hepatocyte isolation, the time when Cyp2e1 expression is still preserved.

In the next step, the classical extrapolation strategy based on AUC and Cmax has been studied.

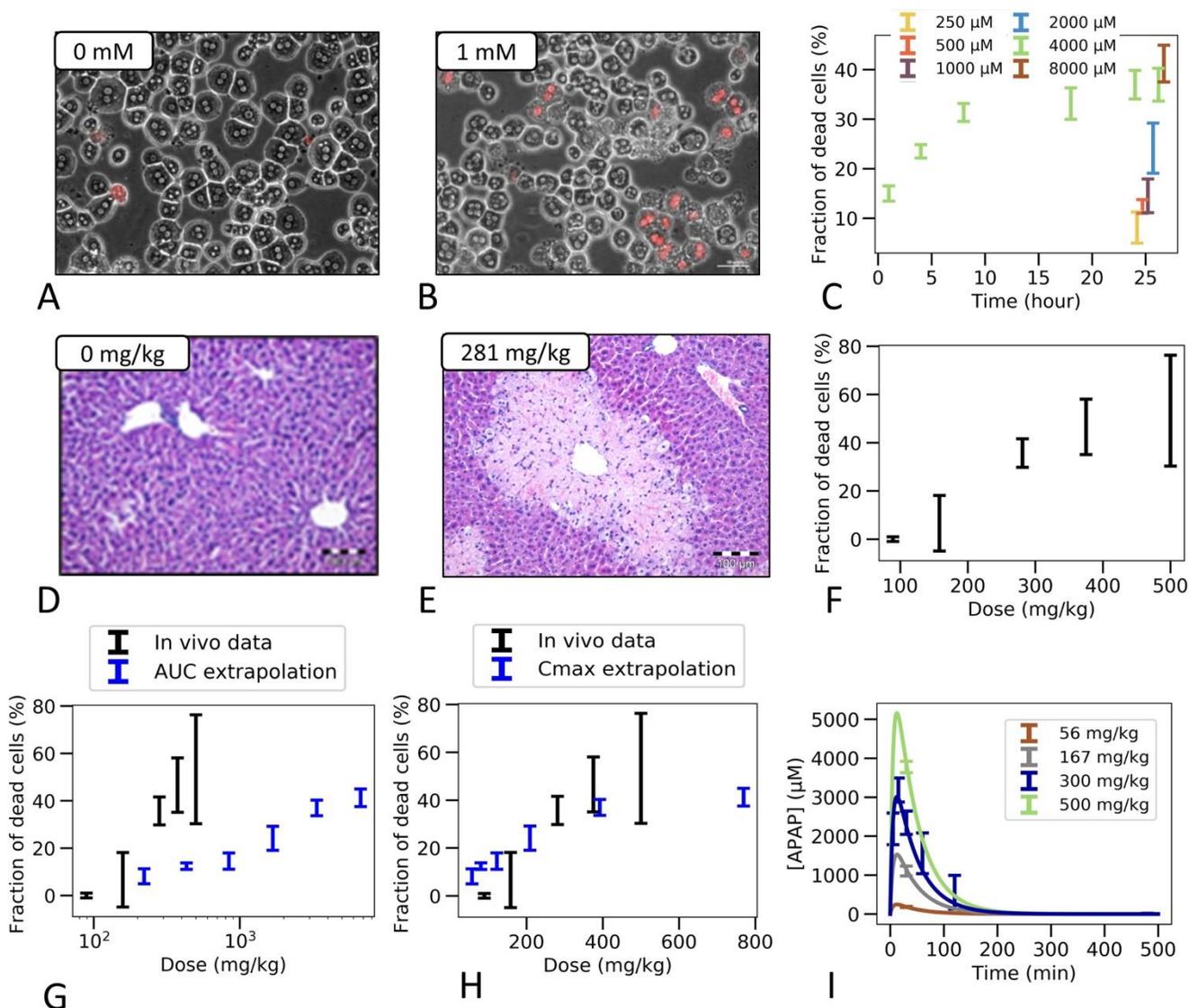

**Figure 2: Experimental *in vitro* and *in vivo* observations.** (A, B) Typical *in vitro* patterns of dead hepatocytes in control and after administration of 1 mM APAP. (C) Concentration-dependent hepatotoxicity *in vitro*. (D, E) Typical *in vivo* liver histology in control and after APAP-administration of 281 mg/kg body weight. (F) Dose-dependent hepatotoxicity *in vivo*. (G, H) Predicted (blue) and real

(black) hepatotoxicity from AUC (G) and Cmax values (F) computed from the (I) Pharmaco-dynamics of the APAP blood concentration for various doses (symbols represent data, lines a PK-model). The PK-model permits determination of the maximal blood concentration Cmax and the area under the PK-curve for each dose. (Further results and details such as the experimental settings are in SFigs. 1-3.)

## 3.2 Direct PBPK based *in vitro* to *in vivo* extrapolation: equivalent exposure does not mean equivalent toxicity

For the classical extrapolation scheme (Fig. 1(I)-(V)), based on AUC and Cmax (Figs. 2G, H), first the pharmacokinetics of APAP in the blood plasma was investigated by a combined experimental - modeling strategy (Fig. 1(IV), Fig. 2I).

For determination of the dose-dependent drug concentration at the site of action, four doses (56, 167, 300 and 450 mg APAP/kg b. w.) were administered to male C57BL6/N mice and blood was taken from the portal vein to determine APAP plasma concentrations. The data was collected 30 minutes after injection for all doses, and over additional time points to determine the temporal kinetics for the 300 mg/kg dose. A simple pharmacokinetic (PK) model (Methods, SI) was then fitted to the data and used to simulate concentration time curves for all APAP doses that have been studied for hepatotoxicity *in vivo* (Fig 2I). The concentration (C(t))-curves show a sharp increase after administration of APAP up to a maximum value Cmax that increases with the dose, and a first-order decay after that maximum. The integral under the simulated C(t)-curves corresponds to the AUC. Hence the C(t)-curves permit to attribute a unique Cmax and an AUC value to each dose.

For the AUC approximation, a concentration-equivalent is defined by performing a time integration over the C(t) curve, generally by $AUC_x = \int_0^T C_x(t)dt$ with $t=0$ denoting the injection (administration) time point of the drug, $T$ the time at which the toxicity is measured, $x \in \{vitro, vivo\}$ identifies whether the *in vitro* or *in vivo* values are used. $AUC_x/T$ again has the unit of a concentration. Hence by equivalently equalizing either $AUC_{vitro} = AUC_{vivo}$ or the concentration values defined by $AUC_{vitro}/T = AUC_{vivo}/T$, this procedure allows plotting the fraction of dead cells *in vitro* and *in vivo* for equivalent exposures (Fig 2G). $AUC_{vivo}/T$ denotes the average APAP blood concentration to which the hepatocytes are exposed to during the observation period $T$ *in vivo*, so that the aforementioned approximation assumes that this average determines the hepatotoxicity. The administered *in vitro* APAP concentration corresponds to $C_{vitro}(t=0)$, which is generally a good estimator for $C_{vitro}(t)$ in the time interval $[0, T]$ for $T=24h$ if the volume of the culture medium is much larger than that of the volume of the hepatocyte population i.e., $C_{vitro}(t \leq T) \approx C_{vitro}(0)$.

For equivalent exposures based on AUC the fraction of dead (PI positive) hepatocytes shows a large deviation of the predicted from the measured value *in vivo* (Fig. 2G).

As an alternative to the AUC, the Cmax values of the PK-curve (C(t)) has been approximated with the drug concentrations administered *in vitro* ($C_{vitro} \approx Cmax$; Fig. 1(II, III)).

The intention for this approach is the assumption that the maximum APAP blood concentration *Cmax* is a good estimator for the APAP concentration used in the *in vitro* experiment, $C_{vitro}(0)$, and that this maximum concentration determines the hepatotoxicity. Hence, again each administrated dose could be associated with a dead cell fraction through its corresponding drug concentration value, permitting to assign a hepatotoxicity value to each dose (Fig. 2H).

However, the comparison of dead cells based on Cmax did not result in a good agreement of *in vitro* and *in vivo* data (Fig 2H). One difference is the lower slope of the concentration effect curve *in vitro*. Moreover, the LOAEL occurs at a slightly lower concentration *in vitro* than *in vivo*, which contrasts the AUC-based comparison.

In conclusion, equivalent exposures *in vivo* and *in vitro* based on Cmax and AUC do not lead to equivalent toxicity. A second problem of this approach is the large difference of the results obtained by the AUC-and the Cmax-based extrapolation: a-priori, one cannot know if basing the extrapolation on Cmax or AUC leads to similar predictions, or if one or the other leads to a better prediction. Thus, the 'direct PBPK model extrapolation' fails to predict the correct *in vivo* toxicity. This calls for a refinement of the models used for extrapolation.

### 3.3   Setting up the intra-cellular detoxification (PD) model *in vitro* and *in vivo*

An alternative approach to the above-described direct PK-based extrapolation is a pharmacodynamic model (PD) mimicking the effect of APAP on each cell of a population (Fig 3A.1, Fig. 4A). The processes considered include diffusion of APAP between extracellular space and cytosol, and the key processes of APAP metabolism.

The final *in vivo* PD model consists of an intracellular PD-model module, and a model module specifying how the APAP feeds into the cell (Fig. 3A.1, Fig. 4A).
The intracellular processes considered are the detoxification by UGT and SULT, and the activating metabolism by CYP2E1 and CYP1A2 (Fig. 4A). CYP enzymes lead to production of NAPQI which later binds to proteins. This triggers production of reactive oxygen species (ROS) modulated by JNK. After ROS accumulate to high levels, the cell reaches membrane permeability transition which impacts ATP production eventually leading to cell death (Snawder et al., 1994; Lee et al., 1996; Genter et al., 1998). The processes, which do not have any impact on the ATP levels (downstream processes) are not modeled here in order to reduce the complexity of the model.

Each cell in the model contains the same ODE system. Briefly, for all parameters of the ODE model a range of physiologically relevant values was used based on published data. To account for the cell-to-cell variability the distribution of the enzyme activity parameters in the cell population was assumed to be log-normal with a coefficient of variation (CoVa, ratio of standard deviation to the mean), which is part of the estimated parameters (cf. section 2.10.2).

Not all cells are CYP-positive as confirmed by immunostaining of CYP enzymes which displays zonation around central veins (SFig. 5). After an image processing step, 50.9% of cells were considered CYP2E1 positive and 60.5% of cells are CYP1A2 positive. Cells that are CYP2E1-positive are also CYP1A2-positive, but not necessarily vice-versa. A gradient of activity is assumed from highest CYP activity close to the central vein to lowest CYP activity in the distant region from the central vein. The gradient values were estimated based on the analysis of images stained for the cytochrome P450 enzymes assuming that the gradient of the gray values translate into a CYP-activity gradient. Because the cells considered in the *in vitro* experiments are harvested from the same animals used for the *in vivo* toxicity analysis, we assume the same fraction of cells expressing the gradient. The CYP enzymes' $V_{max}$ -values were thus not subjected to the cell-to-cell variability according to CoVa.

Because the medium volume is much larger than the total volume of all cells together (which corresponds to the *in vitro* reaction volume), the concentration in the medium does not vary significantly. The APAP concentration $C_{vitro}(t)$ in the medium corresponds to a slightly linearly decreasing function that was estimated from experiments as an input function of each cell APAP metabolic model.

A simulation of the *in vitro* setting thus consists in solving the ODEs involved in 3215 cells (number of cells involved in a piece of virtual lobule as described later) with a given initial concentration of APAP in the medium over time. The cells which have an ATP level below 100 µM are considered as dead.

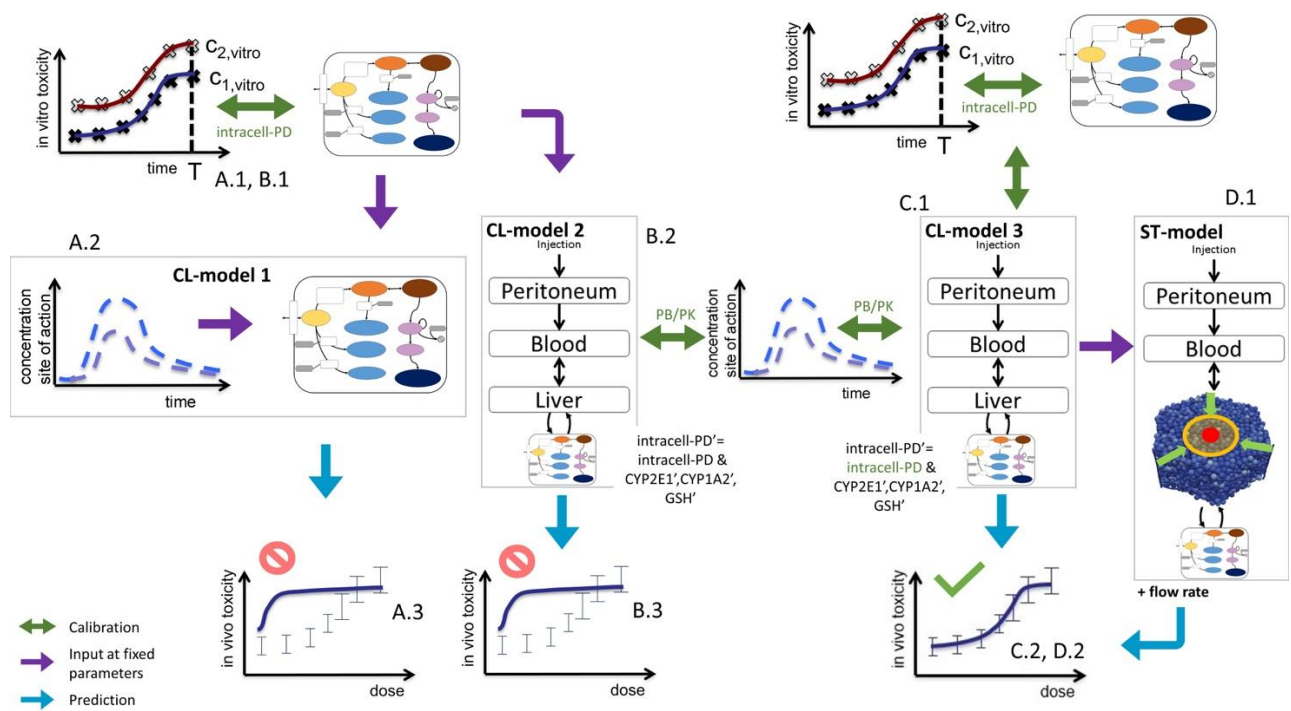

**Figure 3: Sketch of model-based extrapolation strategies**, detailing Fig. 1(VI). For the explanation, see text. Green text indicates those parameters that were fitted in the respective modeling step. The prime in B.2 and C.1 indicate parameters used for the *in vivo* simulation.

The parameters of the PD model were calibrated based on data of the *in vitro* experiments in primary mouse hepatocytes. An excellent agreement between experimental data and PD model could be obtained (Fig. 4B). However, due to the insufficient number of data points and the relatively wide range of physiologically compatible parameter values, no unique parameter set could be identified. Several fits of equivalent agreement quantified by their very similar standard error, were performed using different random seeds, which led to largely different parameter sets. This is expected to impact on the extrapolation to the *in vivo* toxicity.

### 3.4 Stepwise, independent fit *in vitro*, then *in vivo* extrapolation approach fails (CL-model 1)

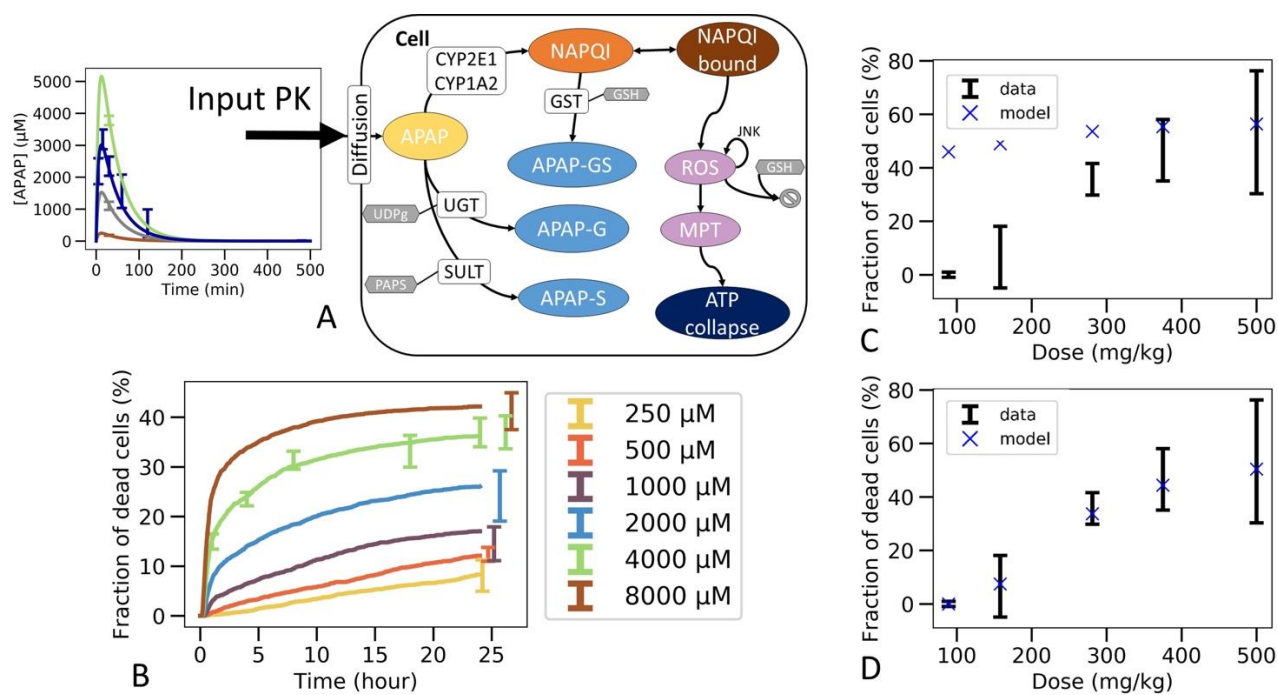

**Figure 4: Non-coupled PK/PD model strategy.** (A) Scheme of the APAP metabolic model where the PK model is plugged as an input, (B) best fit on *in vitro* toxicity data over 40 different random seeds (shown is a particular seed, label 13), extrapolation to *in vivo* fraction of dead cells after correcting CYP enzyme activity and GSH concentration with (C) same permeability *in vitro* and *in vivo* (D) and with lower permeability *in vivo*.

In the next step the parameter sets obtained from the fit of the intracellular model to *in vitro* experimental toxicity data were used to extrapolate to *in vivo* toxicity, considering potential *in vitro / in vivo* differences of the intracellular parameters and in the exposure of hepatocytes to APAP (Fig. 3A).

In a first cycle of model extrapolations, we kept the same parameters for the Cytochrome P450 enzymes and GSH *in vivo* as *in vitro* but extrapolations totally failed (Cellière, 2016). As from those model simulations CYP2E1 and CYP1A2 activities (the two major APAP activating enzymes in mice) have been found to have the strongest impact on the model-based *in vitro – in vivo* extrapolation, they were analyzed in cultivated hepatocytes and freshly isolated hepatocytes, whose activity is known to be similar to those in liver tissue(Godoy et al., 2016) (Godoy et al., 2016) (SFigs. 3, 4). These analyses permitted to quantify the decrease of both enzymes in culture compared to the *in vivo* situation.

Compared to hepatocytes used for the *in vitro* toxicity experiments, the *in vivo* hepatocytes' CYP activities was experimentally found to be lower by a factor of 3.3 for CYP2E1 and 1.8 for CYP1A2 *in vitro* than *in vivo* (SFig. 3), while the GSH concentration *in vivo* was determined to be 2 times lower *in vivo* than *in vitro* (see SFig. 4).

Concerning the difference in APAP exposure *in vivo* and *in vitro*, our first strategy consisted in computing the concentration of APAP in the liver from the previously introduced PK model assuming that the number of APAP-molecules taken up from the blood by the cells during one passage of blood through the liver is consistent with the time course of the number of APAP-molecules in the blood during that passage, and that each hepatocyte *in vivo* is exposed to approximately the same APAP blood concentration. The approach is simple and computationally cheap because one does not need to represent explicitly the uptake of cells from the blood and because the PK-model and the intracellular PD-model can be run independently of each other. The procedure in this case is: (i) fit the intracellular parameters for the PD model with *in vitro* toxicity data (Fig. 3A.1), (ii) adjust the enzyme activities and metabolite concentrations, (iii) fit the extracellular parameters of the PK model on the APAP blood concentration data (Fig. 3A.2). Once the parameters are calibrated, the *in vivo* PD model is simulated for each toxicity dose and compared to data (Fig. 3A3).

This approach significantly overestimated the fraction of dead cells (Fig. 4C) but yielded good extrapolation results if the permeability for APAP *in vivo* was assumed to be significantly lower than *in vitro* (Fig. 4D). This may be justified as *in vivo* APAP in the blood has to pass the space of Disse which contains extracellular matrix, such as collagens. However, the assumption that the uptake of APAP molecules is negligible compared to the number of APAP molecules remaining in the blood turned out to be violated for all parameters. When considering a lower permeability *in vivo*, the hypothesis is verified but the total hepatocytes uptake is way too low and thus not realistic either (less than 1%) (Prescott and Wright, 1973; Dai et al., 2006; Malfatti et al., 2020).

For this reason, we dropped this hypothesis and moved to another approach that explicitly models the interaction between the evolution of the blood concentration and the uptake of the cells.

### 3.5 Stepwise, independent fit *in vitro*, then fit *in vivo* blood concentration extracellular parameters with a coupled PK/PD model approach fails (CL-model 2)

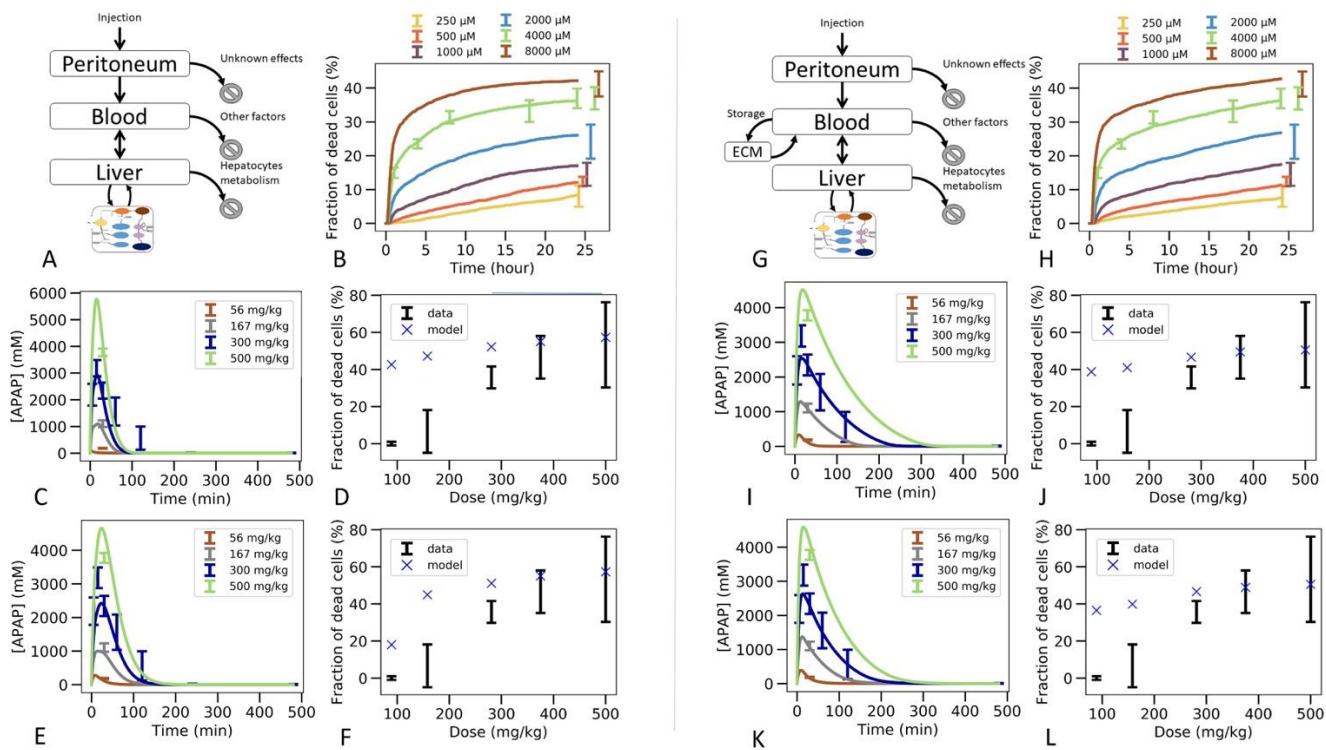

**Figure 5: Two steps coupled PK/PD model strategy.** (A-F) *In vivo* model composed of liver, blood, peritoneal cavity and intracellular PD model. (A) model structure of *in vivo* model, (B) fit of the intracellular PD model on *in vitro* toxicity data for random seed 13 (cf. Fig. 4B), (C) fit of the PK/PD *in vivo* model extracellular parameters on APAP blood *in vivo* concentration data keeping the permeability as its *in vitro* value, (D) extrapolation to *in vivo* toxicity, (E) fit of the PK/PD *in vivo* model extracellular parameters and effective permeability *in vivo* on APAP blood *in vivo* concentration data (E), (F) extrapolation to *in vivo* toxicity. (G-L) *In vivo* model of (A)-(F) extended by an "ECM" compartment representing the space of Disse. (G) Coupled PK/PD model scheme with assumed storage from ECM, (H) fit of the *in vitro*-PD model on *in vitro* toxicity data for random seed 25 that lead to lower permeability, (H) fit of the PK/PD *in vivo* model extracellular parameters on APAP blood *in vivo* concentration data and assumed ECM storage, (I) extrapolation to *in vivo* toxicity, (J), fit of the PK/PD *in vivo* model with additional ECM storage hypothesis, (K) and with extracellular parameters and effective permeability *in vivo* on APAP blood *in vivo* concentration data for random seed 25, (K) extrapolation to *in vivo* toxicity (L).

The second model-based strategy integrated the PK and PD in one single PK/PD model that simultaneously mimics the APAP-PK in the blood, the elimination of APAP from the blood compartment, and the intracellular detoxification pathway reactions (Fig. 5A; Fig. 3B.2). For this purpose, the transport of injected APAP into the blood from the peritoneal cavity as well as its elimination by the liver itself and of all other possible clearance pathways, as for example removal by the kidney, was computed. The elimination by non-liver sources was lumped into one first-order elimination term. The removal of APAP by the liver was modeled by summing the uptake of APAP by all cells of the liver. In practice one sums over the number of cells in a representative slice of the liver lobule of about 10 hepatocytes in height. Then for each cell of the liver lobule slice the set of ODEs representing the intracellular PD model (Fig. 4A) is solved for each time step. This model thus corresponds to a multi-scale model with a well-mixed blood compartment coupled to individual space-free cells (Fig. 5A).

The parameters of the PK/PD are determined in a multi-step procedure (Fig. 3B.1, B.2). (1) Firstly, that set for the parameters of the intracellular model is chosen as a starting parameter set that generated the best fit of the APAP *in vitro*-PD model on *in vitro* toxicity data (Fig. 5B, Fig. 3B.1). (2) Secondly, the activities of CYP2E1 and CYP1A2, as the one of GSH were replaced according to the factors determined in section 2.3. (3) Thirdly, the so modified intracellular model was integrated with the blood and peritoneal cavity compartments to fit the extracellular parameters such that the PK data could be captured (Fig. 5C, Fig. 3B.2). An extra constraint was added to represent elimination by other factors than liver to represent 15% of the total bioavailable dose (Prescott and Wright, 1973; Dai et al., 2006; Malfatti et al., 2020)

Using this second strategy, the fit to the APAP blood concentration (Fig. 5C) misses out the lowest concentration data and still deviates markedly from the profile over time for the 300 mg/kg dose. Moreover, the APAP concentration decreased too quickly with time compared to the data in Fig. 5C. Finally, and most importantly, the fraction of dead cells is largely overpredicted for small APAP doses (Fig. 5D). We hypothesized that the permeability *in vivo* that is determined by both the hepatocyte membrane and the crossing of the space of Disse, may be smaller than *in vitro* (where only the hepatocyte membrane needs to be traversed), which would lower the APAP amount passing into the hepatocyte per time unit, so may reduce cell death in particular for small doses (McPhail et al., 1993). Permeability *in vivo* can thus be considered as an *effective* permeability.

To test this hypothesis now the extra-cellular parameters including the effective permeability were fitted. While this improved the fit quality it was still not enough to fully capture the blood concentration data (Fig. 5E). In particular, the profile over time is closer to the data but the concentration for the lowest concentration data is missed out. Interestingly, the permeability re-fitted led to a lower value of the predicted *in vivo* toxicity, despite the fraction of dead cells *in vivo* was still overestimated (Fig. 5F). (Notice, that refitting of the permeability to the *in vivo* toxicity data after fitting of all other parameters as in Fig. 5C would modify the fit to the PK-data, hence provides no option.)

Since the simulated blood concentration overall fit to the PK data was not great, neither in Fig. 5C, nor in Fig. 5E, a storage mechanism was introduced representing the hypothesis that APAP is transported with a delay into the hepatocytes as it would be the case if it is first stored in the space of Disse (e.g. by adhering to the ECM) and then released from the ECM for transport into the hepatocyte with a delay (Fig. 5G). As lowering the permeability improved the fit quality at the previous step, we chose here the set of intracellular parameters fitted on *in vivo* data for which the permeability was the lowest to test whether this could eliminate the overprediction of *in vivo* toxicity at small doses (Fig. 5H). The storage mechanism improves significantly the capturing of the blood profile over time (Fig. 5I). The lowest concentration data is also captured accurately (Fig. 5I). However, even these two steps, (1) adding a delay in the transport of APAP from the blood into the hepatocyte, (2) choosing the parameter set with the smallest permeability, did not lead to a better toxicity prediction (Fig. 5J).

In a last attempt, letting the permeability *in vivo* be a fit parameter (as in the transition from Fig. 5C to Fig. 5I), slightly improves the fit to the *in vivo* APAP blood concentration (Fig. 5K) but the model could still not predict the *in vivo* fraction of dead cells (Fig. 5L). This is expected as we intentionally chose a random seed parameter set for which the *in vitro* permeability was already low.

In conclusion, fitting the *in vitro* data in a first step (Fig. 3B.1), then replacing *in vitro* cytochrome P450-and GSH activities in that model by their *in vivo* counterparts, and finally fitting a model integrating the so adapted intracellular model with a blood pharmacokinetics (PK) model to the PK data in a last step (Fig. 3B.2), failed to yield reasonable predictions of the *in vivo* toxicity (Fig. 3B.3).

Hence, we asked the question if the integrated PK/PD model if it would be fitted to both the *in vitro* and PK data in a single step, could yield a valid *in vivo* prediction.

### 3.6 Simultaneous parameter determination *in vitro* and *in vivo* blood data extrapolation by a PK/PD model coupled approach succeeds (CL-model 3)

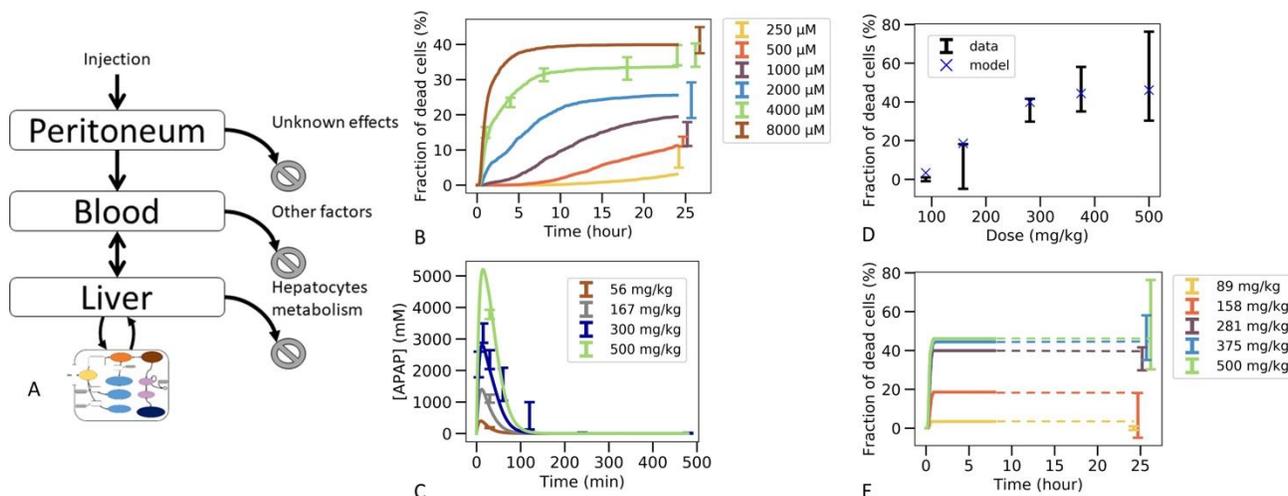

**Figure 6: Simultaneous fit to *in vitro* toxicity data and *in vivo* pharmacokinetic data strategy.** (A) Scheme of the model, (B) fit to *in vitro* toxicity data, (C) fit to *in vivo* pharmacokinetic data, (D) extrapolation to *in vivo* toxicity data at 24h after damage, (E) extrapolation to *in vivo* toxicity data

over time. Simulations were ran only until 480min and since the fraction of dead cells saturates, the final result at 24h is extrapolated from there (illustrated by dotted lines).

In this last fitting strategy (Fig. 3C.1, C.2; Fig. 1(VI-A, B)), the parameters of the intracellular APAP detoxification model were fitted to the *in vitro* toxicity data, and in the same step i.e., simultaneously, the extracellular parameters of the full model, composed of the intracellular *in vivo* model that is obtained by modifying the enzyme activities of CYP2E1, CYP1A2 and glutathione in the *in vitro* model (that is fitted in the same step) according to the experimentally measured factors and the *in vivo* compartments, are fitted to the *in vivo* blood (PK) concentration data (Fig. 3C.1, C.2). The fitness function to minimize was the sum of both fitness function previously used for both data sets individually. For simplicity the storage mechanism of Fig. 5E was not included. The model is able to capture well the data *in vitro* (Fig. 6B) along with *in vivo* blood concentration data (Fig. 6C). When extrapolating to *in vivo*, the model was able to predict the *in vivo* toxicity data with sufficient accuracy (Fig. 6D; Fig. 3C.3). However, for a different random generator value during the fit procedure (for example, starting the fit procedure from different points in the parameter space) an overestimation of the damage at small doses was observed (SFig. 6D), even though the quality of fit to the *in vitro* and PK data did not change (SFig. 6B, C). Nevertheless, that extrapolation still performed better than in the other computational model-based extrapolation strategies.

These results indicate that simultaneously fitting the intracellular parameters to the *in vitro* toxicity and extracellular parameters within a model integrating the intracellular model and the relevant body compartments to the PK data, is able to generate a very good agreement to that data on one hand, and a valid extrapolation to the *in vivo* toxicity data on the other hand. This means that the mechanisms of the model considered here are enough to explain the data available. However, redoing the fit also yielded parameters sets for which the *in vivo* hepatotoxicity prediction failed to be accurate enough although *in vitro* and PK data were well captured. This indicates that in order to take a final decision on whether the studied set of mechanisms is de facto explaining the hepatotoxicity further information is necessary. This can for example be through more narrow parameter ranges or additional *in vitro* toxicity measurements, either determining the hepatotoxicity *in vitro* for additional doses of APAP or measuring at more time points, or directly measuring parameters of the intracellular detoxification model displayed in Fig. 3A. In order to assist identification of those parameters that may be prioritized by future experiments we performed a sensitivity analysis by increasing and decreasing the values for the intracellular parameter values by up to two orders of magnitude, and studying the impact of these variations on the *in vitro* hepatotoxicity (SFig. 8). We find that the non-sensitive parameters, that are not critical and hence may be down-prioritized in direct measurements are: permeability, maximum velocity of CYP2E1 enzyme, Michaelis constant of CYP2E1 enzyme, Michaelis constant of GST enzyme with GSH compound and Hill factor for the ROS production reaction. In a final step we studied whether varying those parameters would change the *in vivo* toxicity in the model, which was not the case (SFig. 9). All other parameters had a significant impact on the *in vitro* toxicity so are expected to equivalently impact the *in vivo* toxicity.

In the liver lobule, the chemical reaction volumes are given by the hepatocytes in a specific spatial organization due to liver lobule zonation. Downstream hepatocytes see lower APAP concentrations that upstream hepatocytes. In order to study in how far this spatial organization may modify the detoxification of APAP, finally the effect of microarchitecture on the APAP detoxification was studied within a spatial-temporal micro-architecture liver lobule model.

### 3.7 Considering lobular microarchitecture by a multi-level virtual liver lobule (ST-model)

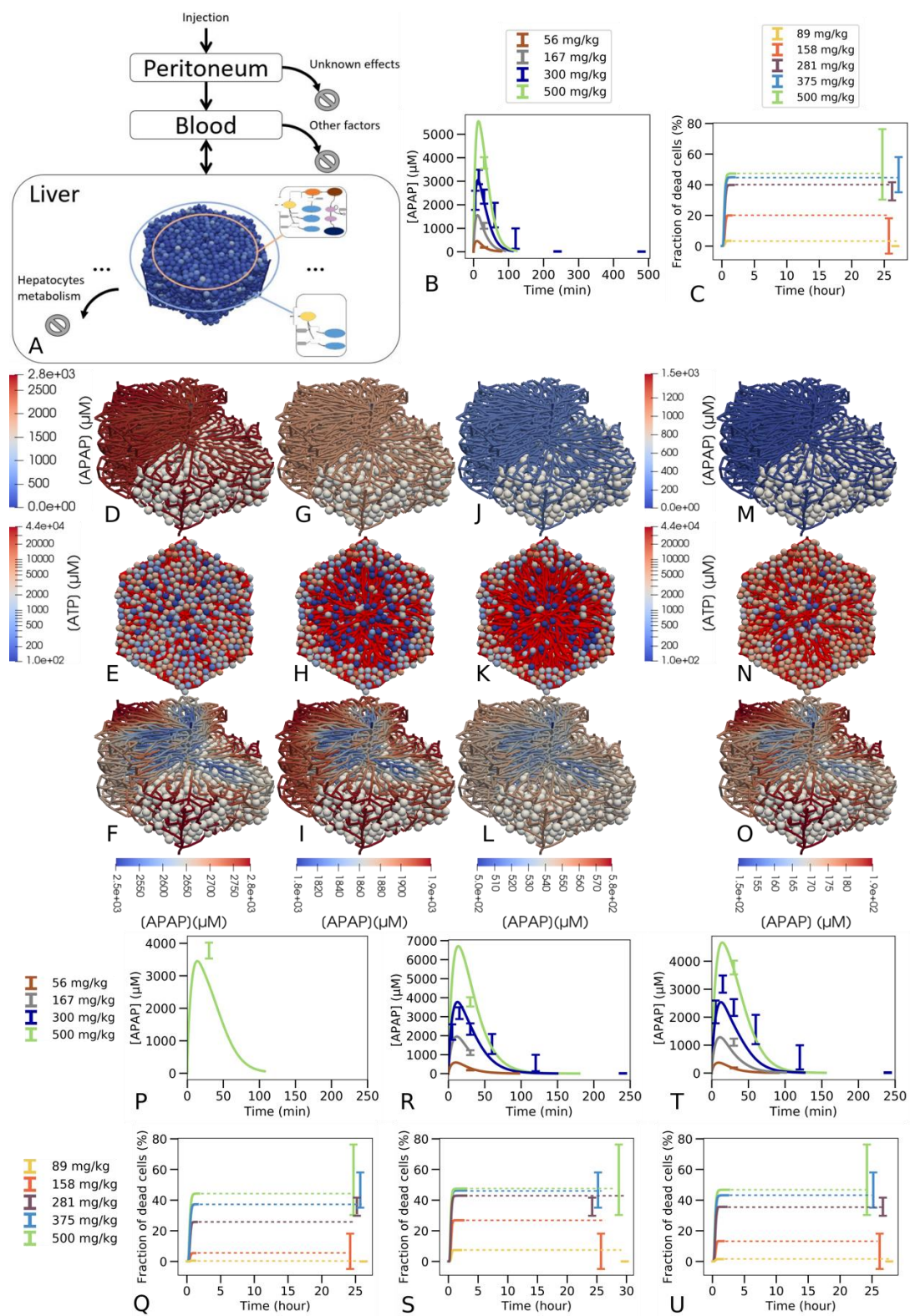

**Figure 7: Multiscale model strategy.** (A) Multiscale model involving a spatial-temporal liver lobule coupled to blood and peritoneum compartments, (B) APAP concentration in blood, (C) fraction of dead cells over time (dotted line is extrapolation after stationary state was reached). (D, G, J, M; F, I, L, O): typical APAP-induced injury scenario in a quasi 3D-liver lobule showing cells as white spheres and the sinusoidal network as system of pipes with their APAP concentration. For the upper

left half of the lobule no cells are shown. In (D, G, J, M) the APAP concentration scale is chosen to compare the lobules among each other, in F, I, L, O it is chosen finer to display concentration changes inside each individual lobule. (E, H, K, N) each displays a liver lobule from above with the spheres representing hepatocytes, colored according to their ATP concentration, while a network of red pipes display the sinusoids. (D) 3D lobule for a 281 mg/kg dose and (E) equivalent 2D representation 15 minutes after injection, (F) the same lobule with minimum APAP concentration value set at 2500µM (see scalebar), (G-H-I) the lobule 30 minutes after injection, minimum concentration value set at 1800µM, (J-K-L) 1 hour after injection with minimum value set at 500µM, (M-N-O) 1 hour after injection for a 158mg/kg dose with minimum value set at 150µM. For D, F, H, J color maps the APAP concentration in the blood vasculature, for E, G, I, K color maps the intracellular ATP concentration are given on a logarithm scale. Note that the peri-central density of living cells drops from E (15mins after APAP injection) to K (1h after APAP injection) so that the sinusoidal network becomes more visible. Due to the smaller APAP dose, there are less surviving cells in K than in N (both 1h after injection of APAP). The results shown in (A-O) are for a volume flow rate of Q=2.685mL/min. (P-U): Blood pharmaco-kinetics and hepatotoxicity prediction for $Q_{liver} = 5.34\ mL/min$ (P, Q), $1.8\ mL/min$ (R, S) and $3.57mL/min$ (T, U).

Finally, we studied whether taking into account the spatial organization of hepatocytes and sinusoids in the liver lobule at otherwise the same parameters as in the well-mixed CL-model 3, one would obtain the same or similar results for the APAP-pharmacokinetics and the *in vivo* hepatotoxicity prediction. To study this question, a multi-level hepatic lobule model was created that considers the complex lobular microarchitecture to include detoxification effects emerging from the spatial organization of hepatocytes within the liver lobule (cf. Fig.1(VI-A, C), Fig. 3D). Such an approach has the fundamental advantage, that it could directly represent disease-related architectural distortions as they occur in fibrosis, steatosis or cirrhosis (Ghallab et al., 2019b, 2021). E.g., in periportal fibrosis, cytochrome P450-negative hepatocytes are partially replaced by scar tissue (Ghallab et al., 2019a), while in septal fibrosis, fibrotic streaks connect central veins of neighboring lobules hence reduce the number of CYP-positive hepatocytes (Ghallab et al., 2019b).

As a proof of concept, the simulations have been performed in a virtual liver lobule generated by sampling from statistical distributions of geometric parameters defined to characterize 3D lobule architecture in confocal laser scanning micrographs of healthy mice (Hoehme et al., 2010) (cf. Fig.1 (VI-C)). The geometric parameters and the topology of the sinusoidal network constitute a further set of parameters compared to the well-mixed models CL 2, 3. The geometric parameter distributions were obtained by analysis of a few tens of liver lobules that could not be directly used for the simulations as none of them displayed an entire lobule. The virtual lobule permits to construct an entire lobule in (x, y)-direction displaying all portal triads (Fig. 7A). This ensured that flow and pressure boundary conditions could be uniquely determined. The representative lobule model represents each individual hepatocyte within the lobule as well as the sinusoidal (the capillary) network. Blood flow is modeled by Poiseuille flow, APAP transport within the lobule by a partial differential equation, and APAP metabolism by the ODE-based PD model that is solved within each individual hepatocyte (SI). The same parameters for the ODE model were used as obtained in the previous section. Cytochrome P450-enzymes were zonated as experimentally observed (cf. SFig. 5).

This virtual spatial liver lobule replaces the set of independent, well-mixed hepatocytes of a classical (compartment) model (see definition above). The APAP injected intraperitoneally enters the blood compartment, which feeds the portal veins of the liver lobule and is at the same time fed via the central vein of the liver lobule (Fig. 7A). Hence, the concentration at the inlet of the liver lobule is set according to the concentration in the blood compartment, while the blood compartment receives inputs from the peritoneum and the central veins. In this approach, the liver blood flow rate ($Q_{liver}$) had to be included in the model to couple the blood compartment to the liver lobule. Besides the geometry parameters this is another additional parameter compared to the well-mixed model, which was adjusted in the range $Q_{liver} \in [1.8\text{mL/min}, 12\text{mL/min}]$ (Davies and Morris, 1993; Schliess et al., 2014; Godoy et al., 2016) (see also SI).

In the spatial-temporal (ST)-model, APAP enters the periphery of each liver lobule via the blood of the portal vein and the hepatic artery. From there it passes through the sinusoidal network along the hepatocytes until it drains into the central vein. In the ST-model, the periportal hepatocytes do not express CYP2E1 or CYP1A2 and therefore CYP450-mediated hepatotoxicity does not take place in this part of the liver lobule. Nevertheless, glucuronidation and sulfation reactions of APAP take place in the periportal hepatocytes, and the formed phase 2 metabolites can no longer contribute to pericentral toxicity. Therefore, periportal hepatocytes already eliminate a fraction of APAP before it reaches the pericentral region without generating cytotoxicity. This effect is automatically accounted for in the ST-model but lacking in the compartmental PD model that lumped all hepatocytes with the same enzyme equipment into one well-stirred compartment (Figs. 1(VI B, C), 5A, 6A). As a consequence, the actual APAP concentration to which the CYP positive cells are exposed to, is smaller in the ST-model (Fig. 7) than in the well-mixed CL-model 3 (Fig. 6). Also, the zonation of the CYP enzymes is taken into account in space in the ST-model by that the hepatocytes' CYP enzyme activities have been chosen to reproduce the experimental gradient from low values in the CYP negative regions to high values in the pericentral region quantitatively (SFig. 5).

Under the assumption that the liver lobules are arranged in parallel i.e., arranged in a such a way that a volume of blood having passed one lobule drains the central vein to leave the liver without entering another liver lobule, the liver volume flow rate (abbreviated mostly as "flow" hereafter) behaves as $Q_{liver} = \frac{\overline{Q}_{lobule}}{N_{lobule}}$, whereby $\overline{Q}_{lobule}$ denotes the average flow leaving the lobule, $N_{lobule}$ is the number of lobules. Hence for a given liver flow, the flow per lobule depends on the number of lobules. Liver flow passing a sinusoid was approximated by Poiseuille flow, hence is proportional to the pressure gradient times the fourth power of the sinusoidal radius: $Q = vA = \frac{\pi}{8\eta(r)} r^4 \nabla p$. Here, $v$ is the flow speed in the sinusoid, $A \approx \pi r^2$ its cross-sectional area. Hence, for a given network topology, the lobule flow is controlled by the flow speed and the sinusoidal radius. Each of the parameters $Q, v, r$ and $N_{lobule}$ was fixed within the range of its experimental measured values found in literature. If $N_{lobule}$ is chosen as for the CL-model 3, the topology and radius as in Hoehme et. al., (2010), then setting a total input flow of $Q_{lobule,in} = 7.2 \times 10^{-5} mL/min$ at the lobule PVs leads to pressure differences of $P_{PV} - P_{CV} = 113 Pa$, an average velocity of $\overline{v} = 47 \mu m \cdot s^{-1}$ and a liver flow of $Q_{liver} = 5.3 mL/min$, which is in line with data found in literature (see Supplementary for more details). However, this parameter setting leads to an underestimation of the APAP blood concentration and the *in vivo* hepatotoxicity (Fig. 7P, Q).

For the same lobular flow but a smaller number of lobules $N'_{lobule} = N_{lobule}/2$, resulting in $Q_{liver} = 2.7 \ mL/min$ (Fig. 7B, C) or $N'_{lobule} = N_{lobule}/1.5$, resulting in $Q_{liver} = 3.57 \ mL /min$ (Fig. 7T, U), an excellent agreement to the *in vivo* toxicity data can be achieved. This is still in the range of possible parameter values for both, $Q_{liver} \in [1.8 mL/min, 12\text{mL/min}]$, and $\frac{N_{liver}}{N_{lobule}} \approx 2.5 \times 10^4 -$ $8.1 \times 10^4$, respectively. Reducing the liver flow rate to $Q_{liver} = 1.8 \ mL/min$ leads on the other hand to an overestimation of the blood concentration and predicts a too high fraction of dead cells (Fig. 7R-S). The observation that we were able to reproduce the hepatotoxicity *in vivo* with the ST-model using a smaller number of lobules $N_{lobule}$ than in the CL-model 3 prompted us to study the sensitivity of the results with regard to a change of $N_{lobule}$ in CL-model 3 (SFig. 7). In agreement with the effect of $N_{lobule}$ on the results of the ST-model, we found that changing the number of lobules $N_{lobule}$ in the CL-3 model also significantly impacts on both APAP blood concentration and the fraction of dead cells *in vivo*.

If the number of lobules should be kept at the same value as in the CL-model 3, a flow of $Q_{liver} = 2.7 - 3.57 \, mL/min$ and $Q_{lobule,in} = 3.6 \times 10^{-5} - 4.8 \times 10^{-5} \, mL/min$ seems to be required. Such lobular flows can either be obtained by reducing the sinusoidal radius, or the blood flow speed in the sinusoidal network, but a change in the lobular flow rate modifies the APAP transport inside the lobule and the fluxes into the hepatocytes, so it is not guaranteed that the same intracellular parameters can be maintained in this case. Because of the significant run time of the ST-model, we refrained from further systematic parameter search simulations with the ST-model to identify possible other parameter set compatible with $Q_{liver} = 2.7 - 3.57 \, mL/min$ in this work.

The multilevel PD spatial-temporal (ST-)model gives the opportunity to visualize APAP concentrations and toxic consequences three-dimensionally in a virtual liver lobule. Simulation snapshots 24 h after administration of 158 mg and 281 mg APAP/kg b.w (Fig. 7K, N) illustrate the dose-dependency of pericentral cell death and allow a comparison to the histological slides of the corresponding doses as those exemplarily presented in SFig. 2C. The 281mg/kg dose was chosen to illustrate the influence of the overdose (damages for 281mg/kg, 375mg/kg and 500 mg/kg are close, see Fig. 7C) and the 158mg/kg to illustrate moderate damage (for 89mg/kg there is almost no damage, see Fig. 7C). Note that LOAEL (Lowest Observed Adverse Effect Level) from these experiments and simulations may be in the range of 89-158 mg/kg b.w.. Comparison of the visualization of the damage for all doses for which hepatotoxicity data was available is provided in supplementary (SFig. 10). Zooming to a finer APAP-concentration scale shows that the concentration difference of liver lobule exit (CV) and entrance (PVs) is only about 10% of entering concentration (Fig. 7F-I-L 15mins, 30mins and 1h after 281mg/kg APAP dose injection and Fig. 7O at 1h after injection of 158mg/kg APAP injection).

In conclusion the results of Figs. 6 and 7 have shown that the same results were obtained with the well-mixed CL-model 3 and the ST-model, if we permit only one parameter to be different, for which the relation of the simulated liver tissue piece cell number (or volume) and the total liver hepatocyte population size (or liver volume) has been chosen. However, both values, the one chosen in CL-model 3 and the one chosen in ST-model, are within the range of values compatible with published knowledge (cf. Supplementary, sect. 1.4). The ST-model has as additional parameter, the liver flow, which, together with the geometry and topology parameters of the liver lobule, determines the liver lobule flow. It is likely that geometry parameters compatible with the experimental ranges can be found so that the aforementioned relation does not need to be adapted. However, the mismatch, even if it can be balanced by parameter changes inside the experimental justifiable parameter ranges, indicates that the assumption of a perfectly well mixed liver compartment may not be correct. A perfectly well-mixed liver compartment would be expected to correspond to the case where the liver lobule flow is so fast that a perfect mixing of each blood volume inside the liver lobule and the blood outside the liver is guaranteed.

This line of argument is supported by looking at the simplest multi-compartment model that requires the liver volume flow rate as a parameter (Supplementary). That simple model suggests that at least one case in which the assumption of one well-mixed compartment with hepatocytes in that compartment seems appropriate is given if the liver flow rate is so large that the flux into the hepatocytes remains negligible compared to the flux along the sinusoids, which for quick enough degradation of APAP means that the liver flow rate is large compared to the permeability-surface product (that has the same unit as the liver flow rate). In that case, a time scale separation occurs, so that first the APAP concentrations in the extrahepatic and intrahepatic blood compartments equilibrate, until the intrahepatic degradation of APAP simultaneously decreases the APAP blood concentration in both blood compartments.

Plugging in numbers for permeability-surface product and liver volume flow rate shows that the ratio of both, $\frac{PS_{liver}}{Q_{liver}}$ is not small compared to one, instead, it is larger than one (Supplementary section 1.6.5). Comparing this ratio for the entire lobule compartment ($\frac{PS_{liver}}{Q_{liver}} = 26.45$) and for a typical "representative sinusoid" (see Supplemental, sect. 1.6.5) composed of about 10 hepatocytes aligned in a row is about the same, given for a single hepatocyte aligning a sinusoid it is $\frac{PS_H}{Q_{sin}} = 2.56$ and $S_{sin} \approx 10 \times S_H$.

Moreover, varying the permeability by a factor $f$ between 100 and about 0.1 to obtain $P' = f \times P$ within a sensitivity analysis in CL-model 3 (SFig. 8) did not change the hepatotoxicity in vivo, while from a factor of $f \lesssim 0.1$, the hepatotoxicity starts to drop indicating that for $f \lesssim 0.1$, the permeability becomes rate limiting, while above, the intracellular APAP-degrading reactions are rate limiting.

Nevertheless, good extrapolations for both the CL-model-3 and the ST-model could be obtained that out-competed the classical extrapolation strategies based on Cmax or AUC (Fig. 8), if the intracellular model is fitted to the *in vitro* data and the full *in vivo* model simultaneously to the pharmaco-kinetics data. The resulting fraction of dead cells obtained in the *in vivo* model then yields the hepatotoxicity.

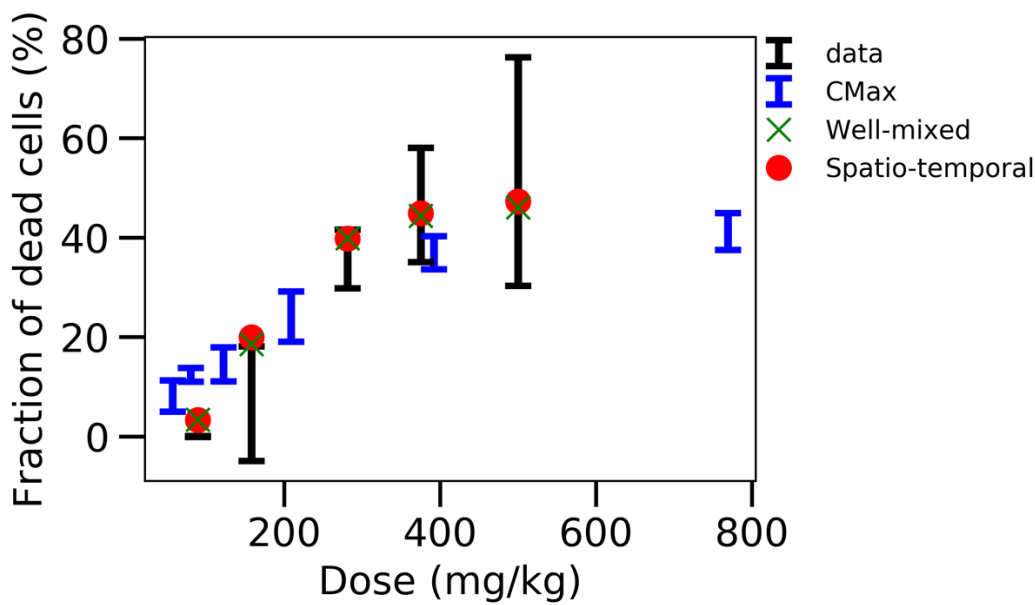

**Figure 8: Comparison of the different strategies.** Displays the fraction of dead cells versus the initial APAP dose *in vivo* for data, CMax strategy, well-mixed model with simultaneous fitting strategy and corresponding ST-model.

## 4    DISCUSSION

An important question in toxicological research is how accurate *in vitro* toxicity data can be extrapolated to the toxicity *in vivo*. Guided by this question, we studied a number of extrapolation strategies for APAP-overdose induced liver injury, starting with the classical extrapolation strategy, followed by mathematical model-based extrapolation strategies varying the complexity of the model and the parameter identification strategy. APAP was chosen as its toxicity mechanism was considered as relatively well understood, so mathematical models could be built based on a scheme of toxicity mechanisms that was believed to largely reflect consensus, in particular involving NAPQI-induced hepatocyte death. In so far, the model results may also inform about in how far the consensus mechanisms permit to explain *in vivo* toxicity.

To permit evaluation of the accuracy of the extrapolation strategy, first hepatotoxicity was experimentally determined over a wide range of concentrations *in vitro* and doses *in vivo*, complemented by measurements of the drug pharmacokinetics (PK) in the blood, activity determinations of the enzymes CYP2E1, CYP1A2, and glutathione concentration, respectively. The *in vitro* measurements use cultivated primary hepatocytes of the same mouse strain that was used *in vivo*.

A number of different approaches (some in different variants) was then studied with regard to their ability to predict hepatotoxicity *in vivo* from *in vitro* hepatotoxicity data:

1. **Str1** (Fig. 1(I-V)): The classical extrapolation scheme based on AUC and Cmax of the blood PK data.
2. **Str2** (Fig. 3-A): A three-step model approach, in which first the parameters of a PD model were determined by fitting the PD model to *in vitro* toxicity data. Then, in a second step, those enzyme activities measured to change *in vitro* from their *in vivo* values were modified in the PD model and the so modified model (CL-model 1) was in a third step used to predict hepatotoxicity *in vivo* by receiving its input by a blood pharmaco-kinetic model for APAP.
3. **Str3** (Fig. 3B): A three-step approach, modifying only step 3 of the previous approach (Str2.) by embedding the modified PD model into a multicompartment *in vivo* model such that the pharmacokinetics was now a model outcome. The intracellular parameters were first fitted on the *in vitro* toxicity data and then the extracellular parameters were fitted on the *in vivo* kinetic data. (CL-model 2)
4. **Str4** (Fig. 1(VI-B), Fig. 3C): A two-step approach using the same model structure as in the previous approach (Str3) but now fitting simultaneously the *in vitro* and *in vivo* model parameters to the *in vitro* hepatotoxicity data and the PK data (CL-model 3). The hepatotoxicity of the model with the so determined parameters corresponds to the *in vivo* hepatotoxicity prediction.
5. **Str5** (Fig. 1(VI-C) Fig. 3D): A spatial-temporal model (ST-model) resolving liver microarchitecture, using the same parameters as determined in strategy (Str4.) plus a liver flow rate parameter

The classical approach (Str1) based on AUC and Cmax of the blood PK-data, complemented by a population PK model, failed to yield an accurate hepatotoxicity prediction. The predicted hepatotoxicity using the AUC is largely deviating from the experimental hepatotoxicity data. Using Cmax the predictions are overall markedly better, but the hepatotoxicity is overestimated for small APAP doses and moderately underestimated for large APAP doses. This points to the difficulty that because the AUC and Cmax-based extrapolation strategies do not yield the same prediction, one would have to know a-priori (which is not the case), which of the two characteristics, AUC or Cmax, should be chosen.

Hence, we studied whether pharmacodynamics (PD) models of the classical well-mixed type (Str2-4) or spatio-temporal (ST; Str5) model resolving micro-architecture may help to overcome the problems of the classical extrapolation strategy (all schematized in Fig 3).

The created model of the PD in a single hepatocyte reflects the consensus mechanisms of APAP hepatotoxicity including conversion of APAP to NAPQI catalyzed by CYP2E1 and CYP1A2, and conversion of APAP by SULT, and UGT. The underlying structure of the intracellular (cell: hepatocyte) PD model was part of each modeling/fitting strategy Str2-Str5.

In strategies (Str2: Fig. 3A.1, Str3: Fig. 3B.1) in a first step the parameters of the intra-cellular PD model were directly fitted to the *in vitro* data taking into account the fraction of hepatocytes in which the cytochrome P450 enzymes CYP2E1 and CYP1A2 are expressed. The intracellular PD model simulates the mechanism of action of APAP, including its activating and inactivating metabolism, generation of reactive oxygen species, mitochondrial toxicity, ATP depletion and cell death.

In a next step, for all strategies, the CYP2E1 and CYP1A2 activities as well as the GSH initial concentrations are of the intracellular PD model were adapted to their *in vivo* – found values to obtain an "*in vivo* intracellular PD-model", before coupling the latter to a PK module (Str2: Fig.4; Str3: Fig. 5; Str. 4: Fig. 6; Str5: Fig. 7). Discrepancies as between Cmax and AUC for the classical extrapolation scheme (Str1) cannot occur anymore. A second main advantage of basing the toxicity extrapolation on a model that uses the same intracellular model structure *in vitro* and *in vivo* is that its mechanistic nature permits to easily incorporate the differences between *in vitro* and *in vivo*. This is particularly helpful when equivalent exposure *in vivo* and *in vitro* does not lead to similar toxicity. This is the case when cells *in vitro* differ, e.g. in their metabolic activities from corresponding cells in a tissue *in vivo*.

The coupling of the intracellular PD model and the PK model has been done in three different ways for Str2, Str. 3, Str4/5 (Str5 is a spatially resolved version of Str4).

For Str2 (Fig. 4, Fig. 3A) the PK model was assumed to remain unaffected by the uptake of APAP by the hepatocytes, which would be a reasonable assumption if the removed APAP per passage of blood through the liver would be consistent with the time course of the APAP blood concentration (CL-model 1). This assumption was violated for most parameter sets that were able to yield a reasonable fit to *in vitro* toxicity and no extrapolation to *in vivo* toxicity data can be achieved. When considering a lower effective permeability *in vivo*, the uptake to cells is too low (< 1%) – and thus not realistic – although excellent extrapolation to *in vivo* toxicity data can be achieved.

To avoid such a violation of self-consistency, in Str3 the PK was mimicked by a mathematical compartment model representing peritoneum, blood and hepatocyte compartment instead of considering the PK-curve as input to the *in vivo* intracellular PD model (CL-model 2; Fig. 3B). I.e., the PK becomes an output of the model integrating physiological body compartments and the intracellular detoxification pathway. In this case, the toxicity at small APAP doses was systematically, partially even largely, overestimated (Fig. 5).

Hence in a further strategy (Str4, Fig. 3C) we fitted the parameters of the intracellular *in vitro* PD model (to the *in vitro* toxicity data) and the parameters of the compartment model simultaneously (to the APAP blood concentration data, Fig. 6), which for several parameter sets yielded subsequently good *in vivo* hepatotoxicity predictions (leading to CL-model 3), however, with varying accuracy. This accuracy – and reproducibility – may be improved by additional sets of experiments as outlined below.

In a final step we studied the influence of tissue micro-architecture on detoxification within a spatial temporal model (ST-model; Str5, Fig. 3D), which readily takes liver zonation into account. For the micro-architecture, a lobule generated by the same lobule generator and using the same lobule parameters as in Hoehme et. al. (2010) has been used. The ST-model required to include the liver flow rate as an additional parameter. The ST-model resembles a virtual tissue twin in that it integrates tissue, cell and molecular scale, and simulates blood flow and fluxes in the blood and from blood to cells, and the APAP metabolism zonated as in a "real" liver lobule. When entering the periportal fraction of the liver lobule in that model, APAP first passes hepatocytes that do not express the enzymes CYP2E1 and CYP1A2, which downstream form hepatotoxic products, but detoxifying UDP-glycuronosyl-transferases and sulfotransferases, which form products that do not harm the cell. Therefore, a fraction of APAP was detoxified by sulfation or glucuronidation before it reached the pericentral fraction of cytochrome P450 expressing hepatocytes. During the sequential passages of blood through each liver lobule from the portal tract towards the central vein the removal of APAP by the cells aligning the sinusoids and not expressing CYP-enzymes looks moderate (APAP drops only by about 10% or less by passing from the portal to the central vein, see Fig. 7F-I-L-O and SFig. 11). However, the ST-model with the lobule architecture and topology parameters fixed as described above, did not permit to readily reproduce the results of the well-mixed model (CL-model 3). Only when, for example, the liver flow rate was chosen in the lower range of the experimentally reported flow values, the same intracellular parameters as in the CL-model 3 led to the same quality of fit of the *in vitro* and pharmaco-kinetic data, which implied the same quality of *in vivo* hepatoxicity prediction, outperforming classical AUC or Cmax schemes (Fig. 8). One way to achieve the liver flow rate to be in that range was to choose the number of liver lobules 30-50% lower compared to the number chosen in the CL-model 3. These values were still within the range obtained from liver cell estimates from the literature.

Another way would have been to reduce the liver lobule flow e.g. by the sinusoidal radius (for which also quite large ranges have been reported) but due the nonlinear dependencies of other ST-model parameters this would have required many computer simulations. Within this paper we refrained from a refit of the geometry parameters, intracellular parameters or a simultaneous refit of the liver flow rate and of the geometry parameters due to the long simulation times of the ST-model of several weeks for one simulation run. A future task will hence be to optimize parametrization of the ST-model either by iterations between the CL-model 3 and the ST-model, or by a speed up of the ST-model.

However, a first analysis indicates that the APAP gradients inside the liver lobule, despite they seem moderate on the first view may still be too large to support the assumption of a perfectly well-mixed liver compartment. The uptake by the cells for the model parameter found too is large to assume that all Cytochrome P450 positive and Cytochrome P450 negative cells "see" the same concentration. This line of argument is supported by that even if the concentration does not drop more than up to 10% between the entrance and the exit of the liver lobule, an original APAP concentration entering the liver for the first time reduces to about 5% of its original concentration after only 30 passages through the liver, which, given one passage by mouse takes about 15s (Debbage et al., 1998), takes 7.5mins.

Further parameters that may play a role is the relative size of the liver blood compartment compared to the extra-liver compartment and the order of the intracellular reaction (which may either amplify or smooth out small differences in extracellular concentrations).

Moreover, further zonation effects as described by (Gebhardt, 1992) may be relevant as well, but to avoid introducing of parameters with too large ranges and given the long simulation times with the ST-model, their integration into the model should be accompanied by quantitative experimental data on the same experimental model. Indeed, results by (Means and Ho, 2019) indicate zonation may have an important impact, although different from our study these authors study APAP detoxification in human where they could not base the zonated reaction rates on quantitative dose-dependent pharmacokinetic and toxicity data. Also, their initial conditions differ from ours, which we had adapted to the experimental setting.

Recently, (Ghallab et al., 2022) have identified a new mechanism that contributes to APAP-induced liver injury *in vivo*. APAP overdose causes a breach of the blood-bile barrier and leakage of bile acids into the sinusoidal blood, which are transported again to the pericentral hepatocytes via the sinusoidal uptake carriers. This results in accumulation of bile acids in the pericentral hepatocytes above toxic thresholds, and thereby aggravates APAP hepatotoxicity. Interestingly, blocking the sinusoidal bile acid uptake carriers strongly ameliorates APAP-induced hepatocyte death (Ghallab et al., 2022). This discovery was made possible using functional intravital imaging techniques which allows imaging at subcellular resolution (Hassan, 2016; Schneider et al., 2021a; Vartak et al., 2021). This novel mechanism was not considered in our modeling approach but can prospectively be integrated into the spatial temporal model. However, during the establishment of the modeling strategies shown in this paper, we had – besides the mechanisms detailed in this work - introduced further black-box mechanisms that act dose-dependent to search for additional potential mechanisms, but even with those a better agreement of data and model than depicted in Figs. 6 and 7 and a higher reproducibility could not be obtained. However, integrating the bile canaliculi network in the ST-model as a direct approach may provide further insight.

In summary, the explanation for the data obtained was the better, the more the structure of the model resembled the real liver micro-architecture and physiology. The final model and fitting strategy permit a reasonable prediction of *in vivo* hepatotoxicity from *in vitro* toxicity measurements. However, the difficulty to reproduce the parameter sets upon multiple fit repetitions indicate that (1) the parameter ranges known for the enzyme activities of all reactions may still be too large, or / and (2) the number of data points may be still too small to ensure that the fitting algorithm always finds the same parameter set for those parameters that are identifiable. Although the obtained extrapolation is reasonable, one cannot rule out the possibility that the consensus mechanisms may be incomplete (Ghallab et. al. 2022). To validate this would require two conditions: (1) Either one could experimentally narrow the ranges of the model parameters, e.g., disprove the parameters we found in Figs. 6 and 7, which yielded a reasonable agreement between model with consensus reaction schemes and experimental data. (2) Or one could find conditions under which the consensus reaction scheme with experimentally further specified parameters (e.g. by additional measurements on enzyme activities etc.) clearly fails. This would require a large set of extra experiments, which may perhaps be better resolved within a community effort, and best within a project with development and refinement of a ST-model and simultaneous experimental measurements at all necessary levels to pinpoint each parameter within narrow ranges. To finally decide on the model, accuracy in measurements is fundamental, otherwise it is not possible to take a final decision on the contribution of the CYP-catalyzed mechanism via NAPQI to hepatotoxicity for APAP overdose. Once this accuracy in the parameters is reached, a sufficiently accurate prediction of NOAEL and LOAEL largely following our fit & modeling strategy Str4 and/or Str5 seems within reach.

With regard to the modeling methodology, we believe that models with spatial representation would prospectively be of major interest as they are naturally able to integrate the bile canaliculi network, and to capture architectural modifications in diseases such as fibrosis or cirrhosis. These are difficult to translate into a coarse-grained model with well-mixed compartments as super-cellular resolution.

## 5 Conflict of Interest

*The authors declare that the research was conducted in the absence of any commercial or financial relationships that could be construed as a potential conflict of interest.*

## 6 Author Contributions

The Author Contributions section is mandatory for all articles, including articles by sole authors. If an appropriate statement is not provided on submission, a standard one will be inserted during the production process. The Author Contributions statement must describe the contributions of individual authors referred to by their initials and, in doing so, all authors agree to be accountable for the content of the work. Please see here for full authorship criteria.

## 7 Funding

DD acknowledges support by the BMBF-grants LiSyM, MSDILI and LiSyM-Cancer as well as by the ANR-project iLITE and EU-project PASSPORT. JD acknowledges support by ANR-project iLite and BMBF-project LiSyM. GC acknowledges support by EU-project PASSPORT. AG is funded by the German Research Foundation (DFG; GH 276/1-1; project no. 457840828).

## 8 Data Availability Statement

The data of this manuscript are available upon request.

## 9 References

Albrecht, W., Kappenberg, F., Brecklinghaus, T., Stoeber, R., Marchan, R., Zhang, M., et al. (2019). Prediction of human drug-induced liver injury (DILI) in relation to oral doses and blood concentrations. *Arch Toxicol* 93, 1609–1637. doi: 10.1007/s00204-019-02492-9.

Bartl, M., Pfaff, M., Ghallab, A., Driesch, D., Henkel, S. G., Hengstler, J. G., et al. (2015). Optimality in the zonation of ammonia detoxification in rodent liver. *Arch Toxicol* 89, 2069–2078. doi: 10.1007/s00204-015-1596-4.

Ben-Shachar, R., Chen, Y., Luo, S., Hartman, C., Reed, M., and Nijhout, H. F. (2012). The biochemistry of acetaminophen hepatotoxicity and rescue: a mathematical model. *Theor Biol Med Model* 9, 55. doi: 10.1186/1742-4682-9-55.

Boissier, N., Drasdo, D., and Vignon-Clementel, I. E. (2021). Simulation of a detoxifying organ function: Focus on hemodynamics modeling and convection-reaction numerical simulation in microcirculatory networks. *Int J Numer Meth Biomed Engng* 37. doi: 10.1002/cnm.3422.

Campos, G., Schmidt-Heck, W., De Smedt, J., Widera, A., Ghallab, A., Pütter, L., et al. (2020). Inflammation-associated suppression of metabolic gene networks in acute and chronic liver disease. *Arch Toxicol* 94, 205–217. doi: 10.1007/s00204-019-02630-3.

Cellière, G. (2016). Multi-scale modeling of hepatic drug toxicity and its consequences on ammonia detoxification.

Cherianidou, A., Seidel, F., Kappenberg, F., Dreser, N., Blum, J., Waldmann, T., et al. (2022). Classification of Developmental Toxicants in a Human iPSC Transcriptomics-Based Test. *Chem. Res. Toxicol.* 35, 760–773. doi: 10.1021/acs.chemrestox.1c00392.

Dai, G., He, L., Chou, N., and Wan, Y.-J. Y. (2006). Acetaminophen Metabolism Does Not Contribute to Gender Difference in Its Hepatotoxicity in Mouse. *Toxicological Sciences* 92, 33–41. doi: 10.1093/toxsci/kfj192.

Davies, B., and Morris, T. (1993). Physiological Parameters in Laboratory Animals and Humans. *Pharmaceutical Research* 10.

Debbage, P. L., Griebel, J., Ried, M., Gneiting, T., DeVries, A., and Hutzler, P. (1998). Lectin Intravital Perfusion Studies in Tumor-bearing Mice: Micrometer-resolution, Wide-area Mapping of Microvascular Labeling, Distinguishing Efficiently and Inefficiently Perfused Microregions in the Tumor. *J Histochem Cytochem.* 46, 627–639. doi: 10.1177/002215549804600508.

Diaz Ochoa, J. G., Bucher, J., Péry, A. R. R., Zaldivar Comenges, J. M., Niklas, J., and Mauch, K. (2012). A multi-scale modeling framework for individualized, spatiotemporal prediction of drug effects and toxicological risk. *Front Pharmacol* 3, 204. doi: 10.3389/fphar.2012.00204.

Drasdo, D., Hoehme, S., and Hengstler, J. G. (2014). How predictive quantitative modelling of tissue organisation can inform liver disease pathogenesis.

Feidt, D. M., Klein, K., Hofmann, U., Riedmaier, S., Knobeloch, D., Thasler, W. E., et al. (2010). Profiling Induction of Cytochrome P450 Enzyme Activity by Statins Using a New Liquid Chromatography-Tandem Mass Spectrometry Cocktail Assay in Human Hepatocytes. *Drug Metab Dispos* 38, 1589–1597. doi: 10.1124/dmd.110.033886.

Franiatte, S., Clarke, R., and Ho, H. (2019). A computational model for hepatotoxicity by coupling drug transport and acetaminophen metabolism equations. *International Journal for Numerical Methods in Biomedical Engineering* 35, e3234. doi: 10.1002/cnm.3234.

Fu, X., Sluka, J. P., Clendenon, S. G., Dunn, K. W., Wang, Z., Klaunig, J. E., et al. (2018). Modeling of xenobiotic transport and metabolism in virtual hepatic lobule models. *PLOS ONE* 13, e0198060. doi: 10.1371/journal.pone.0198060.

Furusawa, C., Suzuki, T., Kashiwagi, A., Yomo, T., and Kaneko, K. (2005). Ubiquity of log-normal distributions in intra-cellular reaction dynamics. *Biophysics (Nagoya-shi)* 1, 25–31. doi: 10.2142/biophysics.1.25.

Gebhardt, R. (1992). Metabolic zonation of the liver: Regulation and implications for liver function. *Pharmacology & Therapeutics* 53, 275–354. doi: 10.1016/0163-7258(92)90055-5.

Genter, M. B., Liang, H.-C., Gu, J., Ding, X., Negishi, M., McKinnon, R. A., et al. (1998). Role of CYP2A5 and 2G1 in Acetaminophen Metabolism and Toxicity in the Olfactory Mucosa of the Cyp1a2(−/−)Mouse. *Biochemical Pharmacology* 55, 1819–1826. doi: 10.1016/S0006-2952(98)00004-5.

Ghallab, A., Hassan, R., Hofmann, U., Friebel, A., Hobloss, Z., Brackhagen, L., et al. (2022). Interruption of bile acid uptake by hepatocytes after acetaminophen overdose ameliorates hepatotoxicity. *Journal of Hepatology* 77, 71–83. doi: 10.1016/j.jhep.2022.01.020.

Ghallab, A., Hofmann, U., Sezgin, S., Vartak, N., Hassan, R., Zaza, A., et al. (2019a). Bile Microinfarcts in Cholestasis Are Initiated by Rupture of the Apical Hepatocyte Membrane and Cause Shunting of Bile to Sinusoidal Blood. *Hepatology* 69, 666–683. doi: 10.1002/hep.30213.

Ghallab, A., Myllys, M., Friebel, A., Duda, J., Edlund, K., Halilbasic, E., et al. (2021). Spatio-Temporal Multiscale Analysis of Western Diet-Fed Mice Reveals a Translationally Relevant Sequence of Events during NAFLD Progression. *Cells* 10, 2516. doi: 10.3390/cells10102516.

Ghallab, A., Myllys, M., H. Holland, C., Zaza, A., Murad, W., Hassan, R., et al. (2019b). Influence of Liver Fibrosis on Lobular Zonation. *Cells* 8, 1556. doi: 10.3390/cells8121556.

Godoy, P., Hewitt, N. J., Albrecht, U., Andersen, M. E., Ansari, N., Bhattacharya, S., et al. (2013). Recent advances in 2D and 3D in vitro systems using primary hepatocytes, alternative hepatocyte sources and non-parenchymal liver cells and their use in investigating mechanisms of hepatotoxicity, cell signaling and ADME. *Arch Toxicol* 87, 1315–1530. doi: 10.1007/s00204-013-1078-5.

Godoy, P., Widera, A., Schmidt-Heck, W., Campos, G., Meyer, C., Cadenas, C., et al. (2016). Gene network activity in cultivated primary hepatocytes is highly similar to diseased mammalian liver tissue. *Arch Toxicol* 90, 2513–2529. doi: 10.1007/s00204-016-1761-4.

Hammad, S., Hoehme, S., Friebel, A., von Recklinghausen, I., Othman, A., Begher-Tibbe, B., et al. (2014). Protocols for staining of bile canalicular and sinusoidal networks of human, mouse and pig livers, three-dimensional reconstruction and quantification of tissue microarchitecture by image processing and analysis. *Arch Toxicol* 88, 1161–1183. doi: 10.1007/s00204-014-1243-5.

Hansen, N. (2006). "The CMA Evolution Strategy: A Comparing Review," in *Towards a New Evolutionary Computation: Advances in the Estimation of Distribution Algorithms* Studies in Fuzziness and Soft Computing., eds. J. A. Lozano, P. Larrañaga, I. Inza, and E. Bengoetxea (Berlin, Heidelberg: Springer), 75–102. doi: 10.1007/3-540-32494-1_4.

Hansen, N., and Ostermeier, A. (1996). Adapting arbitrary normal mutation distributions in evolution strategies: the covariance matrix adaptation. in *Proceedings of IEEE International Conference on Evolutionary Computation* (Nagoya, Japan: IEEE), 312–317. doi: 10.1109/ICEC.1996.542381.

Hassan, R. (2016). Possibilities and limitations of intravital imaging. *EXCLI J* 15, 872–874. doi: 10.17179/excli2016-863.

Heldring, M. M., Shaw, A. H., and Beltman, J. B. (2022). Unraveling the effect of intra- and intercellular processes on acetaminophen-induced liver injury. *npj Syst Biol Appl* 8, 1–16. doi: 10.1038/s41540-022-00238-5.

Ho, H., and Zhang, E. (2020). Virtual Lobule Models Are the Key for Multiscale Biomechanical and Pharmacological Modeling for the Liver. *Frontiers in Physiology* 11. Available at: https://www.frontiersin.org/articles/10.3389/fphys.2020.01061.

Hoehme, S., Brulport, M., Bauer, A., Bedawy, E., Schormann, W., Hermes, M., et al. (2010). Prediction and validation of cell alignment along microvessels as order principle to restore tissue architecture in liver regeneration. *Proc. Natl. Acad. Sci. U.S.A.* 107, 10371–10376. doi: 10.1073/pnas.0909374107.

Holland, C. H., Ramirez Flores, R. O., Myllys, M., Hassan, R., Edlund, K., Hofmann, U., et al. (2022). Transcriptomic Cross-Species Analysis of Chronic Liver Disease Reveals Consistent Regulation Between Humans and Mice. *Hepatology Communications* 6, 161–177. doi: 10.1002/hep4.1797.

Holzhütter, H.-G., Drasdo, D., Preusser, T., Lippert, J., and Henney, A. M. (2012). The virtual liver: a multidisciplinary, multilevel challenge for systems biology. *Wiley Interdiscip Rev Syst Biol Med* 4, 221–235. doi: 10.1002/wsbm.1158.

Jaruchotikamol, A., Takase, A., Itoh, S., Kawasaki, Y., Kondo, S., Sakuma, T., et al. (2009). Alteration of Acetaminophen-induced Cytotoxicity in Mouse Hepatocytes during Primary Culture. *Journal of Health Science* 55, 767–776. doi: 10.1248/jhs.55.767.

Jemnitz, K., Veres, Z., Monostory, K., Kóbori, L., and Vereczkey, L. (2008). Interspecies differences in acetaminophen sensitivity of human, rat, and mouse primary hepatocytes. *Toxicology in Vitro* 22, 961–967. doi: 10.1016/j.tiv.2008.02.001.

Kuepfer, L., Niederalt, C., Wendl, T., Schlender, J., Willmann, S., Lippert, J., et al. (2016). Applied Concepts in PBPK Modeling: How to Build a PBPK/PD Model. *CPT Pharmacometrics Syst Pharmacol* 5, 516–531. doi: 10.1002/psp4.12134.

Lang, T., Klein, K., Fischer, J., Nüssler, A. K., Neuhaus, P., Hofmann, U., et al. (2001). Extensive genetic polymorphism in the human CYP2B6 gene with impact on expression and function in human liver. *Pharmacogenetics and Genomics* 11, 399–415.

Leclerc, E., Hamon, J., Claude, I., Jellali, R., Naudot, M., and Bois, F. (2015). Investigation of acetaminophen toxicity in HepG2/C3a microscale cultures using a system biology model of glutathione depletion. *Cell Biol Toxicol* 31, 173–185. doi: 10.1007/s10565-015-9302-0.

Lee, S. S., Buters, J. T., Pineau, T., Fernandez-Salguero, P., and Gonzalez, F. J. (1996). Role of CYP2E1 in the hepatotoxicity of acetaminophen. *J Biol Chem* 271, 12063–12067. doi: 10.1074/jbc.271.20.12063.

Leist, M., Ghallab, A., Graepel, R., Marchan, R., Hassan, R., Bennekou, S. H., et al. (2017). Adverse outcome pathways: opportunities, limitations and open questions. *Arch Toxicol* 91, 3477–3505. doi: 10.1007/s00204-017-2045-3.

Malfatti, M. A., Kuhn, E. A., Murugesh, D. K., Mendez, M. E., Hum, N., Thissen, J. B., et al. (2020). Manipulation of the Gut Microbiome Alters Acetaminophen Biodisposition in Mice. *Sci Rep* 10, 4571. doi: 10.1038/s41598-020-60982-8.

McGill, M. R., and Jaeschke, H. (2013). METABOLISM AND DISPOSITION OF ACETAMINOPHEN: RECENT ADVANCES IN RELATION TO HEPATOTOXICITY AND DIAGNOSIS. *Pharm Res* 30, 2174–2187. doi: 10.1007/s11095-013-1007-6.

McGill, M. R., Williams, C. D., Xie, Y., Ramachandran, A., and Jaeschke, H. (2012). Acetaminophen-induced liver injury in rats and mice: Comparison of protein adducts, mitochondrial dysfunction, and oxidative stress in the mechanism of toxicity. *Toxicology and Applied Pharmacology* 264, 387–394. doi: 10.1016/j.taap.2012.08.015.

McPhail, M. E., Knowles, R. G., Salter, M., Dawson, J., Burchell, B., and Pooson, C. I. (1993). Uptake of acetaminophen (paracetamol) by isolated rat liver cells. *Biochemical Pharmacology* 45, 1599–1604. doi: 10.1016/0006-2952(93)90300-L.

Means, S. A., and Ho, H. (2019). A spatial-temporal model for zonal hepatotoxicity of acetaminophen. *Drug Metabolism and Pharmacokinetics* 34, 71–77. doi: 10.1016/j.dmpk.2018.09.266.

Meyer, M., Schneckener, S., Ludewig, B., Kuepfer, L., and Lippert, J. (2012). Using Expression Data for Quantification of Active Processes in Physiologically Based Pharmacokinetic Modeling. *Drug Metab Dispos* 40, 892–901. doi: 10.1124/dmd.111.043174.

Olson, K. R., Davarpanah, A. H., Schaefer, E. A., Elias, N., and Misdraji, J. (2017). Case 2-2017 — An 18-Year-Old Woman with Acute Liver Failure. *New England Journal of Medicine* 376, 268–278. doi: 10.1056/NEJMcpc1613467.

Prescott, L. F. (1980). Kinetics and metabolism of paracetamol and phenacetin. *Br J Clin Pharmacol* 10, 291S-298S.

Prescott, L. F., and Wright, N. (1973). The effects of hepatic and renal damage on paracetamol metabolism and excretion following overdosage.: A pharmacokinetic study. *British Journal of Pharmacology* 49, 602–613. doi: 10.1111/j.1476-5381.1973.tb08536.x.

Reddyhoff, D., Ward, J., Williams, D., Regan, S., and Webb, S. (2015). Timescale analysis of a mathematical model of acetaminophen metabolism and toxicity. *Journal of Theoretical Biology* 386, 132–146. doi: 10.1016/j.jtbi.2015.08.021.

Sachinidis, A., Albrecht, W., Nell, P., Cherianidou, A., Hewitt, N. J., Edlund, K., et al. (2019). Road Map for Development of Stem Cell-Based Alternative Test Methods. *Trends in Molecular Medicine* 25, 470–481. doi: 10.1016/j.molmed.2019.04.003.

Schenk, A., Ghallab, A., Hofmann, U., Hassan, R., Schwarz, M., Schuppert, A., et al. (2017). Physiologically-based modelling in mice suggests an aggravated loss of clearance capacity after toxic liver damage. *Sci Rep* 7, 6224. doi: 10.1038/s41598-017-04574-z.

Schliess, F., Hoehme, S., Henkel, S. G., Ghallab, A., Driesch, D., Böttger, J., et al. (2014). Integrated metabolic spatial-temporal model for the prediction of ammonia detoxification during liver damage and regeneration. *Hepatology* 60, 2040–2051. doi: 10.1002/hep.27136.

Schneider, K. M., Candels, L. S., Hov, J. R., Myllys, M., Hassan, R., Schneider, C. V., et al. (2021a). Gut microbiota depletion exacerbates cholestatic liver injury via loss of FXR signalling. *Nat Metab* 3, 1228–1241. doi: 10.1038/s42255-021-00452-1.

Schneider, K. M., Elfers, C., Ghallab, A., Schneider, C. V., Galvez, E. J. C., Mohs, A., et al. (2021b). Intestinal Dysbiosis Amplifies Acetaminophen-Induced Acute Liver Injury. *Cellular and Molecular Gastroenterology and Hepatology* 11, 909–933. doi: 10.1016/j.jcmgh.2020.11.002.

Schwen, L. O., Kuepfer, L., and Preusser, T. (2016). Modeling approaches for hepatic spatial heterogeneity in pharmacokinetic simulations. *Drug Discovery Today: Disease Models* 22, 35–43. doi: 10.1016/j.ddmod.2017.09.002.

Sezgin, S., Hassan, R., Zühlke, S., Kuepfer, L., Hengstler, J. G., Spiteller, M., et al. (2018). Spatio-temporal visualization of the distribution of acetaminophen as well as its metabolites and adducts in mouse livers by MALDI MSI. *Arch Toxicol* 92, 2963–2977. doi: 10.1007/s00204-018-2271-3.

Sigal, A., Milo, R., Cohen, A., Geva-Zatorsky, N., Klein, Y., Liron, Y., et al. (2006). Variability and memory of protein levels in human cells. *Nature* 444, 643–646. doi: 10.1038/nature05316.

Smith, A. K., Petersen, B. K., Ropella, G. E. P., Kennedy, R. C., Kaplowitz, N., Ookhtens, M., et al. (2016). Competing Mechanistic Hypotheses of Acetaminophen-Induced Hepatotoxicity Challenged by Virtual Experiments. *PLOS Computational Biology* 12, e1005253. doi: 10.1371/journal.pcbi.1005253.

Snawder, J. E., Roe, A. L., Benson, R. W., and Roberts, D. W. (1994). Loss of CYP2E1 and CYP1A2 activity as a function of acetaminophen dose: relation to toxicity. *Biochem Biophys Res Commun* 203, 532–539. doi: 10.1006/bbrc.1994.2215.

Spencer, S. L., Gaudet, S., Albeck, J. G., Burke, J. M., and Sorger, P. K. (2009). Non-genetic origins of cell-to-cell variability in TRAIL-induced apoptosis. *Nature* 459, 428–432. doi: 10.1038/nature08012.

Sridharan, K., Ansari, E. A., Mulubwa, M., Raju, A. P., Madhoob, A. A., Jufairi, M. A., et al. (2021). Population pharmacokinetic-pharmacodynamic modeling of acetaminophen in preterm neonates with hemodynamically significant patent ductus arteriosus. *European Journal of Pharmaceutical Sciences* 167, 106023. doi: 10.1016/j.ejps.2021.106023.

Upton, R. N., and Mould, D. R. (2014). Basic concepts in population modeling, simulation, and model-based drug development: part 3-introduction to pharmacodynamic modeling methods. *CPT Pharmacometrics Syst Pharmacol* 3, e88. doi: 10.1038/psp.2013.71.

Vartak, N., Guenther, G., Joly, F., Damle-Vartak, A., Wibbelt, G., Fickel, J., et al. (2021). Intravital Dynamic and Correlative Imaging of Mouse Livers Reveals Diffusion-Dominated Canalicular and Flow-Augmented Ductular Bile Flux. *Hepatology* 73, 1531–1550. doi: 10.1002/hep.31422.

Yu, L. X., and Li, B. V. eds. (2014). *FDA Bioequivalence Standards*. New York, NY: Springer New York doi: 10.1007/978-1-4939-1252-0.

# 10 Figure legends

**Figure 1: *In vitro - in vivo* extrapolation approaches**. *In vitro*, toxicity is measured by exposing cell populations of interest to certain concentrations $C_{vitro}$ of drug (here: APAP) and measuring the fraction of cell death with time (I). The toxicity at a certain time T (often: 24h, or 3d) is considered as the time at which the toxicity is compared (II). A common consensus for extrapolation to *in vivo* is that the *in vitro* concentration value $C_{vitro}$ is identified by the $C_{max}$-value (III) of the corresponding pharmaco-kinetic (PK) curve for the drug in the blood (IV). The PK curve depends on the administered drug dose, so that by identifying $C_{max}$ and $C_{vitro}$, the dose value associated with that $C_{max}$-value can be associated with the toxicity value associated with the corresponding $C_{vitro}$ value (V). (Alternatively, to the $C_{vitro}$ / $C_{max}$ values, the area under the curve (AUC) is used). The model-based strategy mimics drug toxicity in a model by simulating the process of drug detoxification *in vitro* by the cells, whereby the toxicity pathway is explicitly represented in each cell and eventually integrated with a PK model (IV) and a compartment model of the organs of interest (VI). The latter may represent organ microarchitecture, here the liver lobule (VI-C) as repetitive minimal tissue unit or consider a well-mixed compartment (VI-B), both integrating an intracellular PD model (VI-A). Simulations with (VI-B, C) directly yield the *in vivo* toxicity (VII).

**Figure 2: Experimental *in vitro* and *in vivo* observations.** (A, B) Typical *in vitro* patterns of dead hepatocytes in control and after administration of 1 mM APAP. (C) Concentration-dependent hepatotoxicity *in vitro*. (D, E) Typical *in vivo* liver histology in control and after APAP-administration of 281 mg/kg body weight. (F) Dose-dependent hepatotoxicity *in vivo*. (G, H) Predicted (blue) and real (black) hepatotoxicity from AUC (G) and Cmax values (F) computed from the (I) Pharmaco-dynamics of the APAP blood concentration for various doses (symbols represent data, lines a PK-model). The PK-model permits determination of the maximal blood concentration Cmax and the area under the PK-curve for each dose. (Further results and details such as the experimental settings are in SFigs. 1-3.)

**Figure 3: Sketch of model-based extrapolation strategies**, detailing Fig. 1(VI). For the explanation, see text. Green text indicates those parameters that were fitted in the respective modeling step. The prime in B.2 and C.1 indicate parameters used for the *in vivo* simulation.

**Figure 4: Non-coupled PK/PD model strategy.** (A) Scheme of the APAP metabolic model where the PK model is plugged as an input, (B) best fit on *in vitro* toxicity data over 40 different random seeds (shown is a particular seed, label 13), extrapolation to *in vivo* fraction of dead cells after correcting CYP enzyme activity and GSH concentration with (C) same permeability *in vitro* and *in vivo* (D) and with lower permeability *in vivo*.

**Figure 5: Two steps coupled PK/PD model strategy.** (A-F) *In vivo* model composed of liver, blood, peritoneal cavity and intracellular PD model. (A) model structure of *in vivo* model, (B) fit of the intracellular PD model on *in vitro* toxicity data for random seed 13 (cf. Fig. 4B), (C) fit of the PK/PD *in vivo* model extracellular parameters on APAP blood *in vivo* concentration data keeping the permeability as its *in vitro* value, (D) extrapolation to *in vivo* toxicity, (E) fit of the PK/PD *in vivo* model extracellular parameters and effective permeability *in vivo* on APAP blood *in vivo* concentration data (E), (F) extrapolation to *in vivo* toxicity. (G-L) *In vivo* model of (A)-(F) extended by an "ECM" compartment representing the space of Disse. (G) Coupled PK/PD model scheme with assumed storage from ECM, (H) fit of the *in vitro*-PD model on *in vitro* toxicity data for random seed 25 that lead to lower permeability, (H) fit of the PK/PD *in vivo* model extracellular parameters on APAP blood *in vivo* concentration data using random seed 25 and assumed ECM storage, (I) extrapolation to *in vivo* toxicity, (J), fit of the PK/PD *in vivo* model with additional ECM storage hypothesis, (K) and with extracellular parameters and effective permeability *in vivo* on APAP blood *in vivo* concentration data for random seed 25, (K) extrapolation to *in vivo* toxicity (L).

**Figure 6: Simultaneous fit to *in vitro* toxicity data and *in vivo* pharmacokinetic data strategy.** (A) Scheme of the model, (B) fit to *in vitro* toxicity data, (C) fit to *in vivo* pharmacokinetic data, (D)

extrapolation to *in vivo* toxicity data at 24h after damage, (E) extrapolation to *in vivo* toxicity data over time. Simulations were ran only until 480min and since the fraction of dead cells saturates, the final result at 24h is extrapolated from there (illustrated by dotted lines).

**Figure 7: Multiscale model strategy.** (A) Multiscale model involving a spatial-temporal liver lobule coupled to blood and peritoneum compartments, (B) APAP concentration in blood, (C) fraction of dead cells over time (dotted line is extrapolation after stationary state was reached). (D, G, J, M; F, I, L, O): typical APAP-induced injury scenario in a quasi 3D-liver lobule showing cells as white spheres and the sinusoidal network as system of pipes with their APAP concentration. For the upper left half of the lobule no cells are shown. In (D, G, J, M) the APAP concentration scale is chosen to compare the lobules among each other, in F, I, L, O it is chosen finer to display concentration changes inside each individual lobule. (E, H, K, N) each displays a liver lobule from above with the spheres representing hepatocytes, colored according to their ATP concentration, while a network of red pipes display the sinusoids. (D) 3D lobule for a 281 mg/kg dose and (E) equivalent 2D representation 15 minutes after injection, (F) the same lobule with minimum APAP concentration value set at 2500µM (see scalebar), (G-H-I) the lobule 30 minutes after injection, minimum concentration value set at 1800µM, (J-K-L) 1 hour after injection with minimum value set at 500µM, (M-N-O) 1 hour after injection for a 158mg/kg dose with minimum value set at 150µM. For D, F, H, J color maps the APAP concentration in the blood vasculature, for E, G, I, K color maps the intracellular ATP concentration are given on a logarithmic scale. Note that the peri-central density of living cells drops from E (15mins after APAP injection) to K (1h after APAP injection) so that the sinusoidal network becomes more visible. Due to the smaller APAP dose, there are less surviving cells in K than in N (both 1h after injection of APAP). The results shown in (A-O) are for a volume flow rate of Q=2.685mL/min. (P-U): Blood pharmaco-kinetics and hepatotoxicity prediction for $Q_{liver} = 5.34 \, mL/min$ (P, Q), $1.8 \, mL/min$ (R, S) and $3.57 mL/min$ (T, U).

**Figure 8: Comparison of the different strategies.** Displays the fraction of dead cells versus the initial APAP dose *in vivo* for data, CMax strategy, well-mixed model with simultaneous fitting strategy and corresponding ST-model.